%% file: paper.tex
\documentclass[conference]{IEEEtran}
\input{preamble}

\usepackage{amsmath}
\usepackage{hyperref}

\begin{document}

%%%%%%%%%%%%%%%%%%%%%%%%%%%%%%%%%%%%%%%%%%%%%%%%%%%%%%%%%%%%%%%%%%%%%%%%%%%%
% Paper-specific commands {{{

% System name (if applicable)
\newcommand{\name}{$\mathsf{BMS}$\xspace} % middle of sentence
\newcommand{\Name}{$\mathsf{BMS}$\xspace} % start of sentence

% Per-author comment commands (they don't appear if you "make finalized")
% 	Colors: black, blue, brown, cyan, darkgray, gray, green, lightgray, lime
% 	magenta, olive, orange, pink, purple, red, teal, violet, white,	yellow
% \NewCommentType{todo}{red}{TODO}
% \NewCommentType{dml}{red}{dml}
% \NewCommentType{leon}{orange}{leon}

\begin{acronym}
    \acro{DDoS}{Distributed Denial of Service}
    \acro{BMS}{Botnet Monitoring System}
\end{acronym}

% }}}
%%%%%%%%%%%%%%%%%%%%%%%%%%%%%%%%%%%%%%%%%%%%%%%%%%%%%%%%%%%%%%%%%%%%%%%%%%%%

\title{The End of the Canonical IoT Botnet: \\ A Measurement Study of Mirai's Descendants}

%%%%%%%%%%%%%%%%%%%%%%%%%%%%%%%%%%%%%%%%%%%%%%%%%%%%%%%%%%%%%%%%%%%%%%%%%%%%
% Authors {{{

%for single author (just remove % characters)
\author{
{\rm Leon Böck}\\
Telecooperation Lab \\
Technical University of Darmstadt\\
\and
{\rm Valentin Sundermann}\\
Telecooperation Lab \\
Technical University of Darmstadt\and
{\rm Isabella Fusari}\\
George Mason University
\and
{\rm Shankar Karuppayah}\\
National Advanced IPv6 Centre \\
Universiti Sains Malaysia
\and
{\rm Max Mühlhäuser}\\
Telecooperation Lab \\
Technical University of Darmstadt\and
{\rm Dave Levin}\\
University of Maryland
% copy the following lines to add more authors
% \and
% {\rm Name}\\
%Name Institution
} % end author

% }}}
%%%%%%%%%%%%%%%%%%%%%%%%%%%%%%%%%%%%%%%%%%%%%%%%%%%%%%%%%%%%%%%%%%%%%%%%%%%%

\maketitle

\begin{abstract}
\input{abstract}
\end{abstract}

%%%%%%%%%%%%%%%%%%%%%%%%%%%%%%%%%%%%%%%%%%%%%%%%%%%%%%%%%%%%%%%%%%%%%%%%%%%%

\input{new-intro}

\input{new-related}

\input{background}

\input{methodology}

\input{setup}

\input{competition}
\input{similarity}
\input{replyrate}
\input{discussion}

\input{conclusion}

%%%%%%%%%%%%%%%%%%%%%%%%%%%%%%%%%%%%%%%%%%%%%%%%%%%%%%%%%%%%%%%%%%%%%%%%%%%%

\bibliographystyle{IEEEtranS}
\bibliography{conferences,refs}

\end{document}

%% file: preamble.tex
\usepackage{color}
\usepackage[dvipsnames]{xcolor} % IEEE needs this; ACM doesn't
\usepackage{totpages} % IEEE needs this; ACM doesn't
\usepackage{graphicx}
\usepackage[labelformat=simple]{subcaption}
\usepackage{xspace}
\usepackage{multirow}
\usepackage{makecell} % Not sure if we need this
\usepackage[ruled,vlined]{algorithm2e}
\usepackage[nolist]{acronym}
\usepackage{booktabs}
\usepackage{pifont}% http://ctan.org/pkg/pifont

\usepackage{ulem}
\newcommand{\cmark}{\ding{51}}%

\newcommand{\parhead}[1]{\medskip \noindent \textbf{#1}\hskip .1in}
\newcommand{\Parhead}[1]{\noindent \textbf{#1}\hskip .1in}

\ifdefined\isFinalized
\newcommand{\NewCommentType}[3]{}
\else
\newcommand{\NewCommentType}[3]{\expandafter\newcommand\csname #1\endcsname[1]{{\color{#2}{#3: ##1}} }}
\fi

\usepackage[override]{cmtt} % make tt font tighter / less ugly

%% file: abstract.tex
Since the burgeoning days of IoT, Mirai has been established as the
canonical IoT botnet.
Not long after the public release of its code, researchers found many
Mirai variants compete with one another for many of the same
vulnerable hosts.
Over time, the myriad Mirai variants evolved to incorporate unique
vulnerabilities, defenses, and regional concentrations.
In this paper, we ask: have Mirai variants evolved to the point that
they are fundamentally distinct?
We answer this question by measuring two of the most popular Mirai
descendants: Hajime and Mozi.
To actively scan both botnets simultaneously, we developed a robust
measurement infrastructure, \name, and ran it for more than eight
months.
The resulting datasets show that these two popular botnets have
diverged in their evolutions from their common ancestor in multiple
ways: they have virtually no overlapping IP addresses, they exhibit
different behavior to network events such as diurnal rate limiting in
China, and more.
Collectively, our results show that there is no longer one canonical
IoT botnet.
We discuss the implications of this finding for researchers and
practitioners.

%% file: new-intro.tex
\section{Introduction}
\label{sec:intro}

The Mirai botnet~\cite{DBLP:conf/uss/AntonakakisABBB17} heralded a new
era of botnets comprised of many cheap, poorly secured Internet of
Things (IoT) devices.
Mirai demonstrated the potential damaging power of IoT devices;
millions of devices with default passwords, easy-to-exploit
vulnerabilities, and poor software update policies had proliferated the
Internet, and Mirai showed how easy it was to compromise them and
harness their power.
With the public release of its code, many copycat botnets emerged,
making Mirai and its variants the canonical IoT botnet.

Griffioen and Doerr~\cite{DBLP:conf/ccs/GriffioenD20} studied the
maelstrom of many concurrent IoT botnets---all direct descendants of
Mirai---and showed extensive competition amongst them, in that the
botnets targeted many of the same IP addresses.
In essence, the different versions of Mirai were in a ``king of the
hill'' competition to control as much of the IoT landscape as possible. 
Griffioen and Doerr showed that what made Mirai variants succeed was
carving out some unique identity, such as incorporating unique
vulnerabilities, targeting region-specific devices, or employing unique
defenses.

In the three years since their study, Mirai variants have continued to
evolve and carve out their own identities, leading us to ask: what is
the current state of Mirai variants?
Do they still resemble their ancestors, or has their search for unique
identities resulted in highly distinct botnets?

To answer these questions, we performed comprehensive, months-long
measurements of two of the most popular Mirai variants: Hajime and Mozi.
We developed a high-performance measurement infrastructure called Botnet Monitoring System
(\name), and used it to measure both botnets simultaneously.
\Name actively crawls both botnets in their entirety, allowing us to
precisely reason about how much overlap the botnets have, how they are
affected by external network events, and so on.
The resulting active-measurement infrastructure we have built is a
significant contribution, and so we will be making its code publicly
available.

Our measurement results show that---unlike the Mirai variants of the
past---today's most popular Mirai variants have very little in common.
In terms of whom they infect, their IP address overlap is virtually
zero (0.8\%), and their AS- and country-level overlaps are similarly
low.
In terms of their reliability, they also demonstrate fundamental
differences: even after controlling for differences in their
communication protocols, Mozi bots exhibit significantly lower response
rates than Hajime bots.
Even in terms of how they react to external network events, today's IoT
botnets differ: for example, Mozi bots appear to be affected by China's
evening bandwidth throttling far more than Hajime bots~\cite{DBLP:conf/sigmetrics/ZhuM0QEHD20}.

Collectively, these findings demonstrate that differences in
implementation, vulnerability selection, and regional targeting have
resulted in two highly distinct botnets, despite having a common
ancestor.
As a result, we conclude that \emph{there is no longer one canonical
IoT botnet}, and that understanding the behaviors and potential
defenses of IoT botnets requires studying more than one in isolation.

\parhead{Contributions} This paper makes the following contributions:

\begin{itemize}

\item We present \name, a measurement infrastructure that actively
	scans multiple botnets concurrently.

\item We apply \name to two of the most popular IoT botnets, Hajime and
	Mozi, and collect data from both simultaneously for over eight
	months.

\item Analyzing the data from our longitudinal study, we find that 
	there is very little overlap between these two major Mirai
		variants in terms of infected IP addresses. Furthermore, we find differences
		in lifetimes, churn, and response ratio.

\item We find that the AS of a bot infection can affect its lifetime, churn and even response ratio.

\item Collectively, our results show that there is no longer one
	canonical IoT botnet; we discuss the ramifications of this finding
	on future IoT measurement efforts.

\item We will be making \name's source code publicly available.

\end{itemize}

%% file: new-related.tex
\section{Related Work}
\label{sec:related}

The central question behind our work is, to put it simply: how much do today's
IoT botnets differ from one another?
As such, the most closely related work to ours are various papers that have
studied multiple IoT botnets simultaneously.
Griffioen and Doerr~\cite{DBLP:conf/ccs/GriffioenD20} studied variants of the
Mirai botnet soon after its initial release.
They used passive techniques---including honeypots and network telescope
data---to collect attempted attacks from live Mirai (and its variants') bots.
Their results showed considerable overlap between the Mirai variants: they
targeted many of the same IP addresses and geographic locales, and barring
several exceptions had similar behaviors.
Griffioen and Doerr found that the Mirai variants that withstood attack from
other variants did so by distinguishing themselves, either by adopting new
vulnerabilities, targeting specific geographic locations, or employing novel
defenses (e.g., closing ports that other variants did not).
While the botnets demonstrated some differences, they were all still by and
large clearly within the Mirai botnet family.
Their work largely inspires our own, as we seek to understand three years after
their original study on whether IoT botnets continued to evolve and, if so, how
their differences have increased.
Additionally, we differ in terms of how we measure the botnets; rather than
take a passive approach (their measurement infrastructure did not actively seek
out bots, but purely let them come to them), we take an active approach, with
the aim of identifying and communicating with \emph{every} bot in both of our
botnets of study.

In their recent blogpost~\cite{netlabP2Pstatus}, Netlab360 reported general
statistics on the size and location of five active peer-to-peer botnets, including
the two botnets we study: Hajime and Mozi.
Our results confirm theirs in terms of size and locations, but our study goes
considerably further to directly compare the botnets' response rates, reactions
to external network events, as well as how much their individual targets
overlap.

In addition to these IoT-centric studies, there has been other work that
measured multiple non-IoT botnets simultaneously.
Various studies of multiple IRC botnets~\cite{DBLP:conf/esorics/FreilingHW05,
DBLP:conf/imc/RajabZMT06, DBLP:conf/nsdi/RajabZMT07, DBLP:conf/ndss/DagonZL06}
and Windows botnets~\cite{DBLP:conf/nsdi/ZhuangDSWOT08,DBLP:conf/cns/HaasKMMF16, DBLP:conf/cns/HaasKMMF16,
DBLP:conf/sp/RossowAWSPDB13, DBLP:conf/IEEEares/KaruppayahBGMMF17,
DBLP:conf/icc/KaruppayahVHMF16, DBLP:conf/imc/AndriesseRB15, DBLP:conf/nsdi/HolzSDBF08, walowdac, DBLP:conf/ccs/KangCLTKNWSHDK09,DBLP:conf/ccs/Stone-GrossCCGSKKV09,DBLP:conf/nsdi/KanichLEVS08}.
Considerably more is known about these classes of botnets, as they have been
active far longer than IoT botnets.
The IRC botnet studies demonstrated similar characteristics as those found by
Griffioen and Doerr~\cite{DBLP:conf/ccs/GriffioenD20} in the early Mirai
variants; that is, there is considerable competition among many
different IRC botnets.
As we will demonstrate, this characteristic no longer holds amongst two of the
most popular Mirai variants.

Finally, there have been extensive measurements of IoT botnets done mostly in
isolation of one another~\cite{DBLP:conf/ndss/CetinGAKITTYE19,DBLP:conf/uss/AntonakakisABBB17,DBLP:conf/ndss/HerwigHHRL19,botnet2012internet}.
Each of these studies provides, in its own right, important insight into the
inner workings of how these botnets operate and how to defend against them.
The main goal of our work differs in that we seek to understand the differences
between the botnets themselves.
We had originally hoped that we could perform such a study by simply
systematizing these various papers.
However, we determined that performing a direct comparison requires
simultaneous measurements in a controlled fashion, thus motivating the creation
of our new measurement infrastructure which we present in
Section~\ref{sec:methodology}.
Moreover, we introduce a new measurement that, to our knowledge, has not been
presented in prior IoT botnet studies: the bots' response rate.
We find surprising differences between the response rates of Hajime and Mozi
bots.

%% file: background.tex
\section{Review of Hajime and Mozi}
\label{sec:background}

In this section, we review details about the Hajime and Mozi botnets that are relevant to our study.
We focus on the two most relevant aspects concerning this paper: the devices targeted by each botnet and the peer-to-peer communication protocols of each botnet.
For additional information, we refer to existing work for Hajime~\cite{DBLP:conf/ndss/HerwigHHRL19,netlabHajime,radwareHajime} and Mozi~\cite{netlab2019, netlab2021, nozomiMozi, microsoftMozi}.

\subsection{Targeted Devices}

Both Hajime and Mozi build upon the original Mirai telnet brute-forcing source code~\cite{DBLP:conf/ccs/GriffioenD20,DBLP:conf/ndss/HerwigHHRL19, upytcs}. 
Hence, both use the original attack functionality of Mirai based on common username and password combinations for open telnet ports.
%
%https://techgenix.com/gafgyt-botnet-mirai-code/ 
%https://threatpost.com/gafgyt-botnet-ddos-mirai/165424/
% Best: https://www.uptycs.com/blog/mirai-code-re-use-in-gafgyt
%
However, Hajime and Mozi each implement several n-day exploits to infect devices through other means than weak telnet passwords.
Table~\ref{tab:targets} provides an overview of all vulnerabilities Hajime and Mozi use.
Overall, Hajime uses 10 and Mozi 15 vulnerabilities to infect devices. 
Out of those, the two botnets only share four vulnerabilities, including brute forcing telnet passwords.
The low overlap in exploits is interesting, as the implemented vulnerabilities were publicly known for many years and could have been adopted by the other botnet.
We speculate that botmasters avoid overlap in exploits to avoid competition and get exclusive access to vulnerable devices. 
Griffioen and Doerr~\cite{DBLP:conf/ccs/GriffioenD20} have reported previously, that the use of unique username and password combinations gave an advantage to competing Mirai variants.

\input{vulnerability-table}

\subsection{Peer-to-peer Protocols and Monitoring}

A significant evolution over their ancestors is that Hajime and Mozi each implement a peer-to-peer protocol as their command and control channel. 
Peer-to-peer command and control is considered one of the most resilient against take-down efforts~\cite{DBLP:conf/sp/RossowAWSPDB13}.
Its resilience is rooted in each bot being able to disseminate botmaster commands, rather than through a server acting as a single point of failure.
Since bots retrieve their commands from any other bot, they need to communicate directly with each other.
Reverse engineering peer-to-peer protocols enable us to impersonate bots and infiltrate the botnet to identify and track infected devices~\cite{DBLP:conf/sp/RossowAWSPDB13}.
Botnet tracking commonly employs \emph{crawlers} and \emph{sensors}, or a combination of the two~\cite{DBLP:conf/sp/RossowAWSPDB13}.
Crawlers use the peer discovery mechanism of the peer-to-peer protocol to discover and track bots actively.
Sensors are passive entities in the peer-to-peer botnet, waiting for bots to connect to them during the bots' peer discovery.
Hajime and Mozi are parasitic botnets that built their command and control on top of the BitTorrent DHT.
Nevertheless, they differ in their actual implementation, which we describe briefly here.
For full details, we refer to the following articles for Hajime~\cite{DBLP:conf/ndss/HerwigHHRL19,netlabHajime,radwareHajime,rapidityHajime} and Mozi~\cite{netlab2019, netlab2021, nozomiMozi, microsoftMozi}.

\subsubsection{Hajime Protocol}

Hajime bots mimic regular peers in the BitTorrent DHT and generate a unique 20-byte BitTorrent node ID.
They join the network like any other benign peer by bootstrapping through \emph{well-known} peers.
To retrieve configuration files, Hajime uses 20-byte SHA-1 identifiers, called \emph{infohash}.
However, they deviate from the BitTorrent default to derive a \emph{infohash}.
Hajime's approach to deriving the identifier is to generate a SHA-1 hash of the configuration and the current day's timestamp.
To find a payload in the DHT bots perform a lookup of the computed \emph{infohash}, which returns a list of \emph{seeders}, from which the payload can be downloaded.
After the successful download of the latest configuration, bots then start to announce themselves as seeders for the downloaded file.
For the actual download, bots use a custom application layer protocol built on top of the uTorrent Transport Protocol (uTP).
During this download, bots exchange a public key, which we later use as a unique ID for Hajime bots.
The main reason for choosing this public key over a Hajime's bot BitTorent ID is, that it allows us to rule out misidentifying benign peers as bots.

\subsubsection{Mozi Protocol}

Similar to Hajime, Mozi bots are also participants in the BitTorrent DHT, but only implement a subset of the BitTorrent messages~\cite{netlab2019}.
Furthermore, Mozi implements its own ID generation approach, which is also used to distinguish Mozi bots from benign peers. 
The hard-coded configuration of Mozi bots states that a Mozi ID starts with six bytes of ``\texttt{0x38}'', which we will refer to as the \emph{Mozi prefix}.
Furthermore, the latest configuration file specifies the Mozi prefix to start with eight bytes of ``\texttt{0x38}''.
The remaining bytes are generated from a linear congruent generator, similar to glibc's random implementation.
Furthermore, the random number generator is seeded with the current timestamp, parent process ID (\texttt{ppid}) and process ID (\texttt{pid}) as follows: \texttt{srand(time(0)} $\oplus$ \texttt{(ppid} $\oplus$ \texttt{pid))}.
This is important, as we observe many ID collisions for Mozi bots.
These are likely caused by devices of the same type having a similar state, i.e., \texttt{ppid}, \texttt{pid}, and default time, upon reboot.

Another difference between Hajime and Mozi is the exchange of config files.
While Hajime largely emulates BitTorrent's approach to discovering and downloading files, Mozi goes an entirely different route.
Instead of using the DHT to look up an infohash, bots deviate from the BitTorrent protocol if they receive messages from peers that have the Mozi prefix in their ID.
Specifically, upon receiving a \emph{find node} request, there is a probability of one-third, that a bot will reply with a configuration file instead of a \emph{find node} reply.

\subsection{Persistence}

The original Mirai botnet did not persist on infected devices.
Therefore, a simple reboot would disinfect the device.
Many variants, including Mozi, started to implement persistence mechanisms to maintain a hold of infected devices. 
Mozi achieves persistence on devices by copying itself into common autostart folders such as /etc/rcS.d and /etc/init.d.
While this effectively allows Mozi to maintain control over a device, turning off the device still causes the malware to lose its state.
Upon startup, it then re-infects itself and generates a new ID and reconnects to the DHT as if it wasn't infected before.
It is important to note that Hajime does not implement any persistence mechanisms and relies on re-infecting devices if they reboot.

%% file: vulnerability-table.tex
% Sorted by botnet
\begin{table}[]
    \begin{tabular}{@{}llccr@{}}
    \toprule
    Vulnerability    & Devices     & Hajime     & Mozi      & Source                        \\ \midrule                 
    Telnet Passwords & Various     & \cmark     & \cmark    & \cite{netlab2019, nozomiMozi, microsoftMozi} \\
	CVE-2016-10372   & Routers (Eir D1000)     & \cmark     & \cmark    & \cite{netlab2019,DBLP:conf/ndss/HerwigHHRL19}             \\
	CVE-2018-10561   & Routers (Dasan GPON)     & \cmark     & \cmark    & \cite{netlab2019, DBLP:conf/ndss/HerwigHHRL19}            \\
	CVE-2018-10562   & Routers (Dasan GPON)    & \cmark     & \cmark    & \cite{netlab2019, DBLP:conf/ndss/HerwigHHRL19}            \\ \hline
	CVE-2017-20149   & Routers (MikroTik)     & \cmark     &           & \cite{netlabHajime, DBLP:conf/ndss/HerwigHHRL19}                \\
	CVE-2015-0966    & Modems (ARRIS)      & \cmark     &           & \cite{DBLP:conf/ndss/HerwigHHRL19, radwareHajime}     \\
	CVE-2015-7289    & Modems (ARRIS)      & \cmark     &           & \cite{DBLP:conf/ndss/HerwigHHRL19}                        \\
	CVE-2017-822\{1..5\}    & IP Cameras (Wificam)  & \cmark     &           & \cite{netlabHajime}             \\
	CVE-2015-4464    & DVRs (KGuard)        & \cmark     &           & \cite{netlabHajime, DBLP:conf/ndss/HerwigHHRL19}                \\
    CVE-2017-17562   & CCTV DVRs   & \cmark     &           & \cite{DBLP:conf/ndss/HerwigHHRL19}                        \\ \hline
    FTP Passwords    & Various     &            & \cmark    & \cite{netlab2021}                    \\
    SSH Passwords    & Various     &            & \cmark    & \cite{netlab2021}                    \\
	CVE-2014-8361    & Routers (Various)     &            & \cmark    & \cite{netlab2019}                    \\
	CVE-2013-7471    & Routers (D-Link)     &            & \cmark    & \cite{netlab2019}                    \\
	CVE-2015-2051    & Routers (D-Link)     &            & \cmark    & \cite{netlab2019}                    \\
	CVE-2017-17215   & Routers (Huawei)     &            & \cmark    & \cite{netlab2019}                    \\
	CVE-2016-6277    & Routers (Netgear)     &            & \cmark    & \cite{netlab2019}                    \\
	CVE-2014-2321    & Modems (ZTE)      &            & \cmark    & \cite{microsoftMozi}                 \\
    CVE-2008-4873    & SPBOARD     &            & \cmark    & \cite{netlab2019}                    \\
	CVE-2016-20016   & DVRs (MVPower)        &            & \cmark    & \cite{netlab2019}                    \\
    NO CVE           & CCTV DVRs   &            & \cmark    & \cite{netlab2019}                    \\ \bottomrule
    \end{tabular}
    \caption{Vulnerabilities target by Hajime and Mozi.}
    \label{tab:targets}
\end{table}

%% file: methodology.tex
\section{\Name Measurement Infrastructure Design}
\label{sec:methodology}

\begin{figure*}[h]
    \centering
    \includegraphics[scale=0.5]{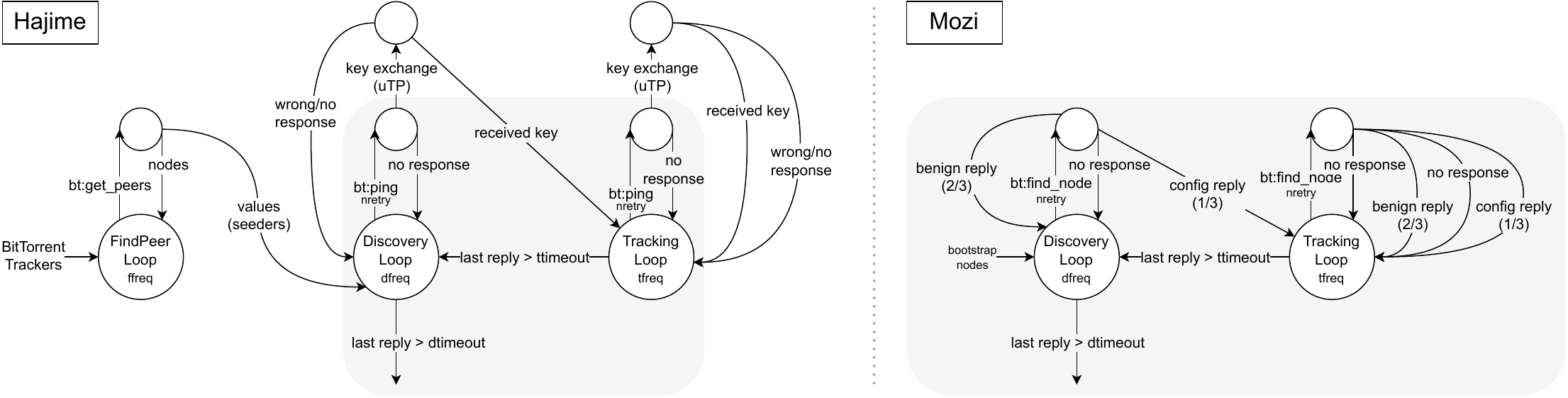}
    \caption{State Diagram for our Hajime and Mozi Crawlers.}
    \label{fig:bms_loops}
\end{figure*}

In this section, we describe our measurement infrastructure, \name.
The central goal of \name is to permit simultaneous active crawls of 
multiple IoT botnets.
As prior work has demonstrated~\cite{DBLP:conf/ndss/HerwigHHRL19},
scanning Hajime alone can be a significant undertaking, requiring many
measurement probes to be sent within a strict timeline.
Mozi requires even more effort, as its bots are distributed throughout
the BitTorrent DHT (see Section~\ref{sec:background}).

\Name's design takes advantage of the commonalities between Hajime and
Mozi, permitting code reuse when possible (thereby reducing effort and
ensuring a more accurate direct comparison), while also allowing
for extensible plugins necessary to meet the botnets' unique behaviors.
The primary components of \name are (1)~the \emph{base crawler}, an
extensible crawler implementation for both UDP and TCP-based botnet
protocols, and (2)~a storage and analysis backend.
The crawlers can be deployed in a distributed manner, and represent the actual data collection endpoints. 
The collected data is then forwarded to the backend to store and process the collected data of all crawlers in a uniform manner.
In the remainder of this section, we describe these components in
detail.

\Name is primarily written in Go, Python and PLpgSQL.
Go is used for performance-critical components such as the backend API and crawlers. 
Python is used primarily for data analysis and enrichment purposes.
PLpgSQL is used to generate tables and automation tasks in our TimescaleDB\footnote{https://www.timescale.com/}
Overall the project consists of 54,203 lines of code.

\subsection{Backend} % {{{
Our main reason for having a single backend is to simplify the analysis and comparison of botnet tracking data.
To that end, \name implements a minimal data format that covers information commonly available in any botnet and from any source, e.g., crawlers, sensors, honeypots, or network telescopes.
This data encompasses a timestamp for an interaction, the IP address, a port, and optionally an ID of the bot.
Furthermore, we also record if a crawl attempt failed, which is necessary for one of our later analyses. 
Other data, e.g., details of a configuration file can be stored in additional json fields.
The backend also implements features to automatically enhance the dataset with AS and geolocation information from MaxMind's GeoLite2 database\footnote{\Name includes GeoLite2 data created by MaxMind, available from https://www.maxmind.com}.

% }}}

\subsection{Crawler Implementation} % {{{

\Parhead{Crawl Loops}
Traditionally crawlers, e.g., web or peer-to-peer network crawlers, are used to repeatedly collect exhaustive \emph{snapshots} of websites or hosts. 
This is commonly achieved following breadth-first search or depth-first search.
However, large peer-to-peer botnets will change throughout an exhaustive crawl due to bot and IP churn.
One main problem is dealing with unresponsive peers slowing down the overall process.
To address this, we crawl continuously using two separate loops to first identify active bots, and then track them separately from the discovery process.

Figure~\ref{fig:bms_loops} depicts the entire crawling process for Hajime and Mozi. 
Both crawlers include a \emph{discovery loop} to discover new bots and
a \emph{tracking loop} to track all discovered bots continuously.
Each loop can run at a different frequency allowing us to track known bots at a higher frequency (dfreq, tfreq) without being slowed down by unresponsive hosts.
Furthermore, we limit unsolicited traffic to benign hosts assigned a previously malicious IP address.

Once the crawler verifies that an IP address in the discovery loop is an active bot, it is moved to the tracking loop. 
If the bot becomes unresponsive for more than \emph{ttimeout} minutes, it is moved back to the discovery loop.
If the bot remains unresponsive for another \emph{dtimeout} minutes in the discovery loop, it is removed entirely.

As shown in Figure~\ref{fig:bms_loops}, Hajime requires more states than Mozi.
The main reason for this is, that Mozi uses a single message to receive either information about other bots or configuration files.
Furthermore, Hajime requires an additional message exchange to retrieve the uTP key of a Hajime bot.
Nevertheless, large parts of the codebase are shared. 
These parts are highlighted in the figure by a light grey box.

\parhead{Network Congestion}
The responsiveness of a bot depends on two factors: its own availability and the availability of the network.
While TCP has built-in functionality to deal with network congestion, UDP does not.
To address the reliability of UDP, we implemented a retry mechanism for UDP-based protocols.
Specifically, we resend a request up to \emph{nretry} times, each time doubling the waiting time starting from two seconds.
With the default setting of $nretry = 5$, this results in a waiting time of up to 62 seconds.

% }}}

%% file: setup.tex
\section{Experimental Setup}

\begin{figure*}
    \centering
    \begin{subfigure}{0.48\textwidth}
        \includegraphics[scale=0.7]{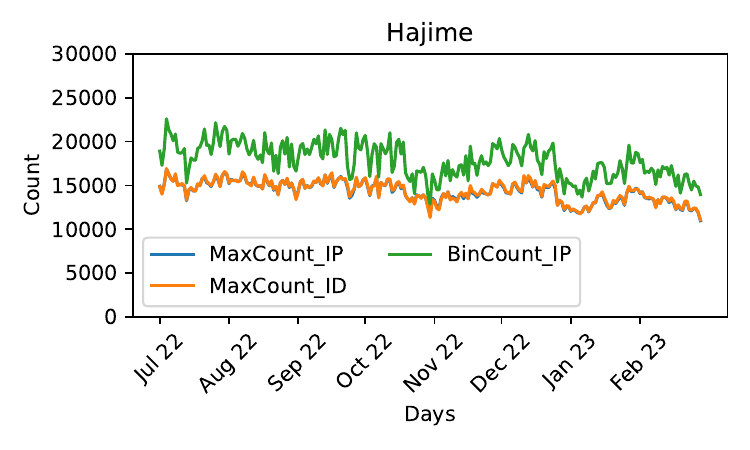}
    \end{subfigure}
    \begin{subfigure}{0.48\textwidth}
        \includegraphics[scale=0.7]{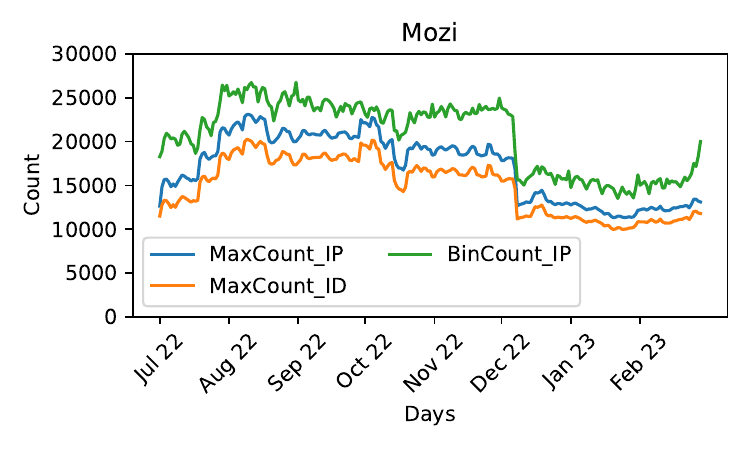}
    \end{subfigure}
    \caption{Botnet sizes throughout the eight-month measurement period. }
    \label{fig:botnet_size}
\end{figure*}

In this section, we present our crawler setup, data post-processing and the description of our final dataset.

\subsection{Crawler Setup}

\begin{table}[]
    \centering
    \begin{tabular}{@{}lllllll@{}}
    \toprule
                & Crawlers & dfreq & tfreq & dtimeout & ttimeout & nretry \\ \midrule
                Hajime      & 4        & 300s & 60s & 900s                                                                & 900s & 5   \\
                Mozi        & 3        & 300s & 60s & 900s                                                                & 900s & 5   \\ \bottomrule
    \end{tabular}
    \caption{Crawler setup and configuration for each botnet. }
    \label{tab:crawler_setup}
\end{table}

Table~\ref{tab:crawler_setup} presents the number and configuration parameters of all our crawlers. 
We initially started with two crawlers for each botnet in our pre-measurements.
However, since Hajime uses the current day's timestamp to compute the infohash of the configuration file, we added two crawlers to look up tomorrow's and yesterday's infohash.
For Mozi we experienced that running all crawlers simultaneously decreased the data retrieved by the two Mozi crawlers.
Consequently, we added a third crawler in a separate location to account for potential losses in our network.
For all other parameters, we chose the same values for each botnet.
Except for the offsite Mozi crawler, all crawlers run in individual VMs, with individual IPs, on the same virtualization server.
Throughout the eight-month measurement, every botnet had at least one crawler running.
Furthermore, with few exceptions, crawlers restarted autonomously after a crash providing stable coverage most of the time.

\subsection{ID Collisions}

As IP addresses are, at best short-term identifiers~\cite{DBLP:conf/ndss/BockLPDM23} we leverage the IDs implemented by both Hajime and Mozi to conduct comparisons of bot behavior.
However, as the generation of Mozi IDs is not perfectly random (c.f. Section~\ref{sec:background}), two bots may have the same ID. 
Duplicate IDs complicate the identification of a single device. 
To address this issue, we increase the uniqueness of Mozi IDs by concatenating the ID and ASN of a bot.
Since the devices targeted by Mozi are mostly routers and other IoT devices, they are unlikely to change location or ASN.
Moreover, a move likely coincides with a loss of power and the generation of a new ID upon reboot.

In addition to the colliding IDs in Mozi, an error in our crawler implementation would sometimes cause an IP address to be recorded with a wrong bot ID.
We addressed this issue by identifying and removing occurrences of this error from our dataset. 

\subsection{Metrics}

In addition to common metrics and statistics, we adapt the terminology and metrics for size estimation proposed by Bock et al.~\cite{DBLP:conf/ndss/BockLPDM23}.
Specifically, we use BinCount and $MaxCount_{AS}$.
BinCount is the intuitive way to count bots, by counting all unique IP addresses observed over a time window $T$.
However, due to IP address reassignments, it is susceptible to overestimation as bots are oftentimes assigned multiple IP addresses, especially for longer time windows.
$MaxCount_{AS}$ addresses this by counting the maximum bots that were active simultaneously in an AS and then summing the maxima of all ASes.
The formal definitions are as follows:

\[\text{BinCount}_T = |S_T|\]

\[\text{MaxCount}_{AS} = \sum_i \max_p |\text{AS}_i(S_T^p)| \]

Where $S_T$ represents the set of IP addresses over some time window $T$ and $p$ denotes (potentially different) times within $T$ where an AS has its maxima. 

\subsection{Dataset Details}

Our dataset spans eight months between July 2022 and February 2023, for which we had at least one crawler running at all times.
Within these eight months, we recorded 2,127,232 IDs and 881,616 IP addresses for the Hajime botnet. 
For the Mozi botnet, we recorded 2,558,652 ID-ASN pairs and 919,518 IP addresses.
Additionally, as these eight-month aggregates are strongly affected by IP churn and changing IDs due to reboots, Figure~\ref{fig:botnet_size} depicts the daily size for Hajime and Mozi.
Overall, Hajime and Mozi are similar in size with their $MaxCount_{AS}$ averaging $14.303$ and $17.125$ bots per day. %\todo{update with actual average}
Similarly, their BinCounts are $17.868k$ and $20.672$, respectively.
A more comprehensive analysis of the botnet's sizes is presented in Section~\ref{sec:comparison}

The differences between IP and ID-based MaxCounts in Mozi are due to the aforementioned ID collisions.
While combining ID and AS allows us to differentiate ID collisions across ASes, we have no reliable way to do the same for ID collisions within an AS.
Consequently, we might underestimate the number of Mozi bots if multiple bots share an ID within the same AS.
Using IP addresses instead of IDs has similar problems, as it can lead to overestimation of the botnet's size~\cite{DBLP:conf/ndss/BockLPDM23}.
Therefore, we chose the more conservative measurement and accept the possibility of underestimating the size of Mozi rather than overestimating it.

%% file: competition.tex
\section{How Similar are the Populations of Hajime and Mozi?}
\label{sec:overlap}

As presented in Section~\ref{sec:background} one of the main evolutions over the original Mirai is, that Hajime and Mozi each implement additional exploits.
Furthermore, only a few of these exploits are shared between Hajime and Mozi.
Therefore, we want to find out how these deviating evolutions affect competition over hosts between Hajime and Mozi.
Griffioen and Doerr have previously observed strong competition of Mirai variants over vulnerable hosts~\cite{DBLP:conf/ccs/GriffioenD20}.
However, they also reported that adding regional-specific username-password combinations led to a single botnet's exclusive infection of hosts.
\subsection{Competition over Vulnerable Hosts}

\begin{figure}[h]
    \centering
    \includegraphics[scale=0.7]{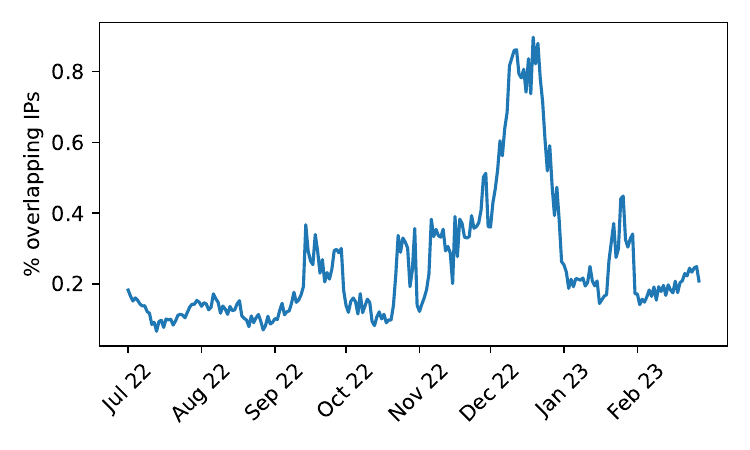}
    \caption{Percentage of overlapping IPs.}
    \label{fig:overlap}
\end{figure}

\begin{figure*}
    \centering
    \includegraphics[scale=0.25]{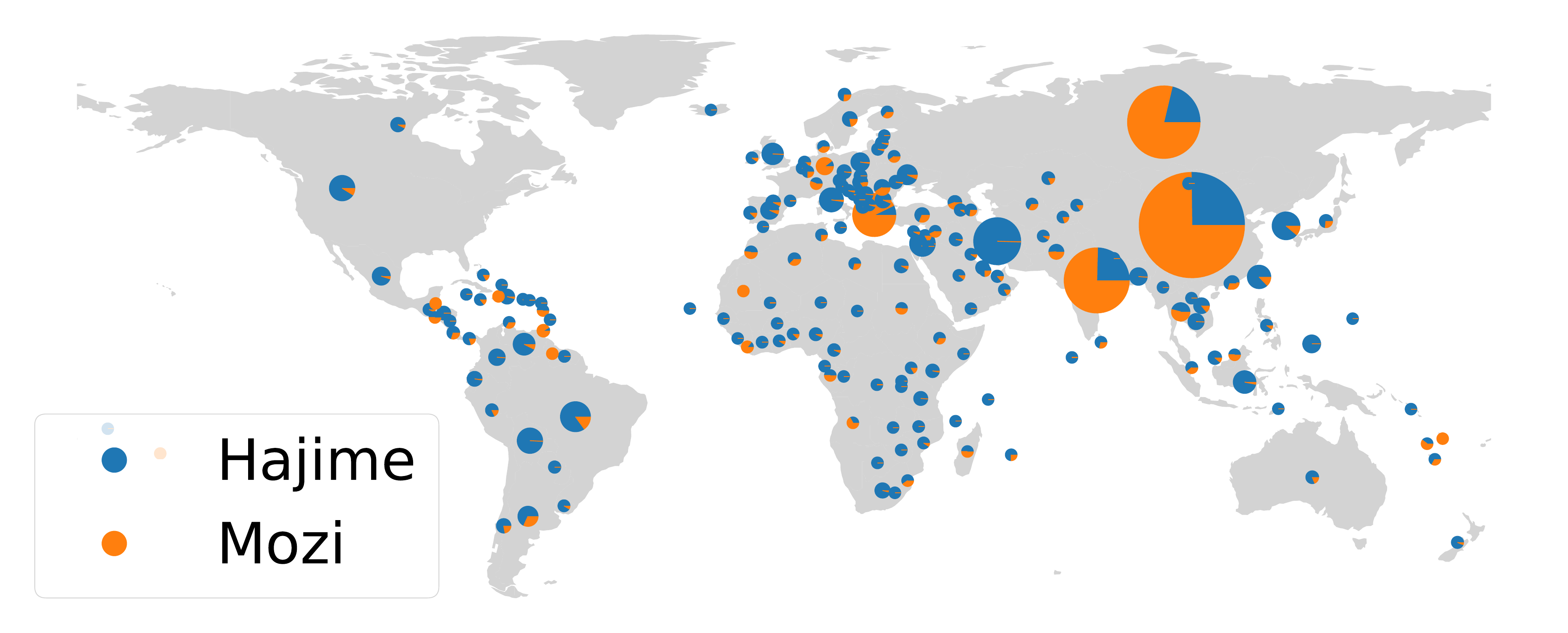}
    \caption{Distribution of global infections for Hajime and Mozi. The size of the circle indicates the number of infections.}
    \label{fig:worldmap}
\end{figure*}

To identify possible competition between the botnets, we analyzed if multiple botnets infect the same IP address.
Since IP addresses change over time~\cite{Padmanabhan16}, we limited our analysis to IP addresses that are infected by at least two botnets in 24 hours.
Figure~\ref{fig:overlap} depicts the daily percentage of IPs overlapping between Hajime and Mozi.
Nearly all of this overlap occurs in three ASes, with rare, single overlaps in other ASes.
Throughout the eight-month measurement, overlap remained below 1\%.
The spike in overlap in December 2022 is related to a general increase in infections of the Mozi botnet in AS 4134.

An average overlap below 0.5\% indicates that the populations of the two botnets are mostly different.
Furthermore, two botnets infecting the same IP within 24 hours does not mean they compete for the same device.
Instead, it could also be due to IP reassignment, or multiple bots sharing a public IP address due to NAT.

To better understand the little overlap we found, we apply two heuristics to distinguish competition from network-related events.
Our heuristic is based on the observation that Hajime and Mozi close vulnerable ports to prevent infection by another botnet. 
Therefore, two botnets cannot infect the same device at the same time.
Thus, the simultaneous activity of two botnets is likely due to two bots sharing a public IP address. 

Our first heuristic aims to identify simultaneous infections based on overlapping activity.
To identify overlaps we search for a sequence of measurements for an IP being infected by $botnet_a \rightarrow botnet_b \rightarrow botnet_a$.
This sequence indicates parallel infections and therefore sharing of the IP address.
Since both botnets prevent simultaneous infection of the same device, it is most likely, that two different bots share a public IP address.

Our second heuristic identifies a possible takeover, i.e., one botnet infecting a device after it was infected by the other.
If the IP is never infected by two botnets simultaneously, and first infected by $botnet_a$ and within six hours by $botnet_b$, we consider it a \emph{possible} takeover of a bot.
This heuristic is likely overestimating the actual takeovers, as IP reassignments could lead to the same sequence of observations.
Based on these two heuristics, we have observed 611 (3.77\%) possible takeovers out of 16,221 IP addresses that had infections from multiple botnets.

Considering that we only observed 611 possible takeovers over eight months, we conclude that there is no competition over hosts between Hajime and Mozi.
Furthermore, what overlap exists, is primarily caused by NAT rather than takeovers between the Hajime and Mozi botnet.
This result differs from Griffioen and Doerr's study of different Mirai variants, which reported substantial amounts of competition and takeovers~\cite{DBLP:conf/ccs/GriffioenD20}.
We attribute this to Mirai using a single attack vector, exploiting similar username password combinations to infect devices, whereas Mozi and Hajime each implement unique exploits in addition to weak telnet passwords.
Our observation, combined with separate exploits, aligns with Griffioen and Doerr's reporting that the AKUMA and MASUTA variants avoided competition by adding region-specific usernames and passwords.

These findings show, that Hajime and Mozi have not only evolved their code base but also started to infect disparate populations.
Another interesting observation is, that original Mirai variants could have been much larger if they had added additional exploits.
This is especially concerning, as Mirai was already responsible for some of the largest volumetric DoS attacks, even threatening major anti-DoS providers~\cite{akamai-ddos}

\subsection{Device Composition}

As we have found no overlap in the populations of Hajime and Mozi, we wanted to investigate if it is indeed related to the unique exploits implemented by each botnet.
For that, we used the Shodan\footnote{www.shodan.io} service to retrieve information about the infected devices.
As we do not have access to a paid plan, we limited the Shodan queries to two days of IP addresses for each botnet.
This results in 22,293 IPs for Hajime and 24,256 IPs for Mozi.
For Hajime, we got a response for 11,444 (51.33\%) IP addresses, whereas we only got a response for 3,314 (13.66\%) of Mozi's IP addresses.
Such a low response rate is to be expected, as Hajime and Mozi close vulnerable ports after infection.
Therefore, scanners like Shodan can only enumerate the ports before a device is infected.
Furthermore, we speculate that the discrepancy in responses is primarily caused by Mozi's persistence mechanism. 
This causes a device to be re-infected with Mozi immediately after reboot, leaving no window for Shodan to scan the bot.

Regardless of the low response rate, we tried to identify devices by the open ports, CVEs reported by Shodan, and regexes for the manufacturer and device identifier of targeted devices.
Most notably, we found that Hajime IPs had 4,631 replies that matched onthe regex (?i)MikroTik, and 1,129 on the regex (?i)cross web server.
These regexes are related to CVE's unique to Hajime (CVE-2017-20149, CVE-2017-17562).
Similarly, for Mozi we found 728 IPs with open FTP ports which are not targeted by Hajime.
While these results are not conclusive, they at least support that Hajime and Mozi are successfully infecting devices with their non-shared exploits.

\subsection{Locations of Infected Devices}

While we could not identify direct competition in the previous section, we want to understand if Hajime and Mozi even infect the same countries and ASes.
For that, we look at the countries that are affected by each botnet.
To obtain a general understanding of global infections Figure~\ref{fig:worldmap} shows the geolocation of bots based on the country of Hajime and Mozi.
Both botnets have infections globally, with Hajime affecting 182 countries and Mozi affecting 130 countries.
The affected countries drop if we filter countries that never exceeded a MaxCount of ten infections to 94 for Hajime and 36 for Mozi.
The discrepancy in affected countries is odd, as both botnets scan most global IP addresses with few exceptions, e.g., known military and government-related IP ranges.
Consequently, Mozi's exploits seem to be more region-specific than Hajime's.

\begin{figure*}
    \centering
    \begin{subfigure}{0.48\textwidth}
        \includegraphics[scale=0.7]{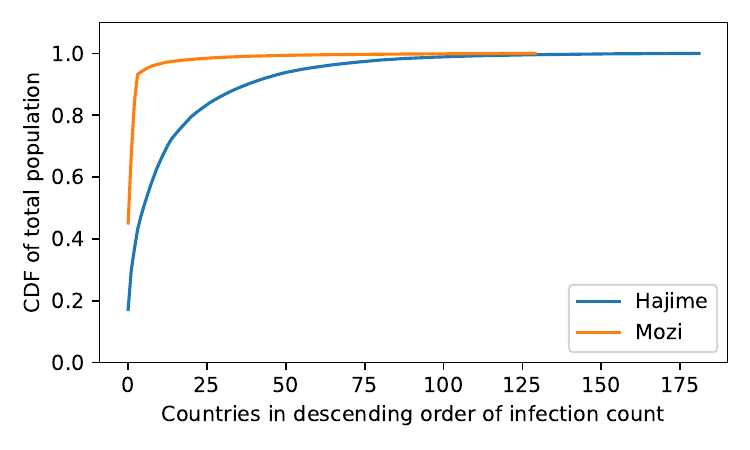}
        % \caption{Mozi botnet size.}
        % \label{fig:mozi_size}
    \end{subfigure}
    \begin{subfigure}{0.48\textwidth}
        \includegraphics[scale=0.7]{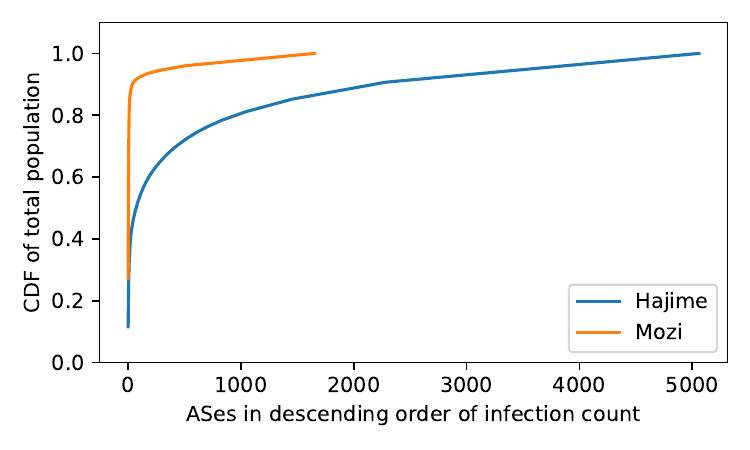}
        % \caption{Hajime botnet size.}
        % \label{fig:hajime_size}
    \end{subfigure}
    \caption{CDFs of infection count by AS and country.}
    \label{fig:infection_cdf}
\end{figure*}

Furthermore, Hajime not only infects more countries, but the number of infected hosts per country is also more extensive.
Tables~\ref{tab:top_10_countries} and~\ref{tab:top_10_AS} provide a more detailed look at the top 10 infected countries and ASes for Hajime and Mozi.
Only five countries and two ASes are in the top 10 of both botnets.
Furthermore, Figure~\ref{fig:infection_cdf} shows the discrepancies in the overall distribution of infections across countries and ASes.
For Mozi, 93.19\% of all infections are in China (10,716), India (4,039), Russia (5,270), and Greece (2,090).
Only eight other countries ever had more than 100 infections.
Contrarily, Hajime-affected countries exceeding 1,000 infections make up only 43.05\%.%only in China (3,610), Iran (2,700), Russia (1,425), and India (1,319), making up 43.05\% of all infections. 
Furthermore, Hajime has 33 countries with at least 100 infections each, whereas Mozi only has eight.

%Countries
% Mozi SUM 23731
% Hajime SUM 21031

\begin{table}[]
    \centering
    \begin{tabular}{@{}llrlr@{}}
    \toprule
         & \multicolumn{2}{l}{Hajime}                 & \multicolumn{2}{l}{Mozi}                   \\ \midrule
    Rank & Country     & \multicolumn{1}{l}{MaxCount} & Country     & \multicolumn{1}{l}{MaxCount} \\
    1    &  \textbf{CN}          & 3610 (17.17\%)     &  \textbf{CN}          & 10716 (45.16\%)                       \\
    2    & IR & 2700 (12.84\%)                        &  \textbf{RU}          & 5270  (22.21\%)                       \\
    3    &  \textbf{RU}          & 1425 (06.78\%)     &  \textbf{IN}          & 4039  (17.02\%)                       \\
    4    &  \textbf{IN}          & 1319 (06.27\%)     & GR                    & 2090  (08.81\%)                       \\
    5    &  \textbf{BR}          & 849  (04.04\%)     & DE                    & 190   (00.80\%)                       \\
    6    &  \textbf{KR}          & 742  (03.53\%)     &  \textbf{BR}          & 150   (00.63\%)                       \\
    7    & PS & 678  (03.22\%)                        & TH                    & 143   (00.60\%)                       \\
    8    & BO & 662  (03.15\%)                        & AR                    & 112   (00.47\%)                       \\
    9    & US & 608  (02.89\%)                        &  \textbf{KR}          & 93    (00.39\%)                       \\
    10   & IT & 568  (02.70\%)                        & TW                    & 69    (00.29\%)                       \\ \bottomrule
    \end{tabular}
    \caption{Top 10 countries for Haime and Mozi based on infection count ($MaxCount_{AS}$). Bold countries are within the top 10 of both botnets.}
    \label{tab:top_10_countries}
\end{table}

% ASes
% Mozi SUM 28783
% Hajime SUM 29743

\begin{table}[]
    \centering
    \begin{tabular}{@{}llrlr@{}}
    \toprule
         & \multicolumn{2}{l}{Hajime} & \multicolumn{2}{l}{Mozi}   \\ \midrule
    Rank & AS              & MaxCount & AS              & MaxCount \\
    1    &  \textbf{4134}             & 3436  (11.55\%)   & 4837   & 7817 (27.18\%)    \\
    2    & 58224  & 1775  (05.97\%)   & 8402   & 4671 (16.24\%)    \\
    3    &  \textbf{9829}             & 965   (03.44\%)   &  \textbf{9829}            & 2858 (09.94\%)    \\
    4    & 4766  & 675   (02.27\%)    & 6799  & 2090 (07.27\%)    \\
    5    & 27839  & 633   (02.13\%)   & \textbf{4134}          & 1918 (06.67\%)    \\
    6    & 8048  & 397   (01.33\%)    & 17622 & 1199 (04.17\%)    \\
    7    & 12975  & 379   (01.27\%)   & 133661 & 1040 (03.62\%)    \\
    8    & 3269  & 354   (01.19\%)    & 17816 & 791  (02.75\%)    \\
    9    & 51407  & 324   (01.09\%)   & 133696 & 533  (01.85\%)    \\
    10   & 31549  & 264   (00.89\%)   & 44257 & 404  (01.40\%)    \\ \bottomrule
    \end{tabular}
    \caption{Top 10 ASes for Haime and Mozi based on infection count ($MaxCount_{AS}$). Bold ASes are within the top 10 of both botnets.}
    \label{tab:top_10_AS}
\end{table}

Based on the results of both country and AS level comparisons of Hajime and Mozi, we see little similarity in location and geographic distribution.
Even though China, India and Russia are among the most affected countries in both botnets, China and India are the two most populous countries worldwide.
Moreover, there are only one Chinese and one Indian AS in the top ten most affected ASes for both botnets.

Furthermore, two primary differences stand out in the distribution of infected devices.
i) The number of countries and ASes with at least ten infections is more than double for Hajime.
ii) Mozi's infections are concentrated in a few locations, with its top ten infected countries and ASes making up 96.38\% and 81.02\% of all infections.
Contrarily, Hajime's top ten infected countries and ASes comprise only 62.58\% and 30.94\%.
These results show that even if the type of devices (IoT) and spreading (worm-like global scanning) of botnets may be similar, their locations may deviate significantly.
We speculate the reason to be rooted in regional limitations of Mozi's exploits, i.e., Mozi's exploits targeting devices that are only used in some regions across the world. 
Herwig et al.~\cite{DBLP:conf/ndss/HerwigHHRL19} have previously observed regional biases with Hajime infections in Brazil being predominantly mipseb, in China mipsel, and arm5 in the USA.
Furthermore, Griffioen et al.~\cite{DBLP:conf/ccs/GriffioenD20} reported regional biases in Mirai variants caused by vendor-specific username and password combinations.

The difference in geographic distribution also impacts the potential value of a botnet.
This affects both the resilience of the botnet and its potential for DDoS and other attacks. 

The primary strength of peer-to-peer botnets is their resilience to takedowns, as they do not have a single point of failure.
Therefore, defenders have to take down every bot to fully dismantle the botnet, or find an exploitable bug to sinkhole the botnet~\cite{DBLP:conf/sp/RossowAWSPDB13}.
However, the extreme regional biases observed for Mozi strongly undermine the resilience provided by its peer-to-peer communication protocol.
Since ISPs control Internet access to all devices in their AS, they could isolate all infections.
This has been done in the past~\cite{DBLP:conf/ndss/CetinGAKITTYE19} and could prove an effective tool to remediate most Mozi infections.
In the case of Mozi, only ten ISPs could effectively take down 82.45\% of Mozi, making it more centralized than its peer-to-peer structure might suggest. 
Consequently, investigating the underlying location distribution of each botnet should be standard practice to uncover weak spots in distributed botnets.

Aside from the reduced resilience, geographic concentration can reduce the effectiveness of attacks.
Using DDoS attacks as an example, a victim could remediate the majority of Mozi's attack traffic by blocking just ten ASes. 
Furthermore, if the victim is located in a region with few Mozi infections, blocking traffic may not even affect legitimate customers and users.

In summary, we found that Hajime and Mozi affect largely different hosts and show little to no competition over vulnerable devices.
Moreover, Mozi infections are more concentrated in a few countries and ASes, whereas Hajime infections are more distributed across the world.
This makes Hajime more resilient to potential remediation efforts and provides benefits for potential attacks.
Therefore, device composition and geographic location are important factors for both botmasters and defenders to consider in the future. 

%% file: similarity.tex
\section{How Comparable are Population Dynamics in Hajime and Mozi?}
\label{sec:comparison}

In the previous section, we established that Hajime and Mozi have largely disjoint populations.
Considering that, we want to investigate how the bot lifetime and bot churn of Hajime and Mozi differ.
For that, we conduct and compare the results of typical botnet measurements of bot lifetimes and bot churn, on an overall and AS level.

\subsection{Bot Lifetimes}

\begin{figure}
    \centering
    \includegraphics[scale=0.7]{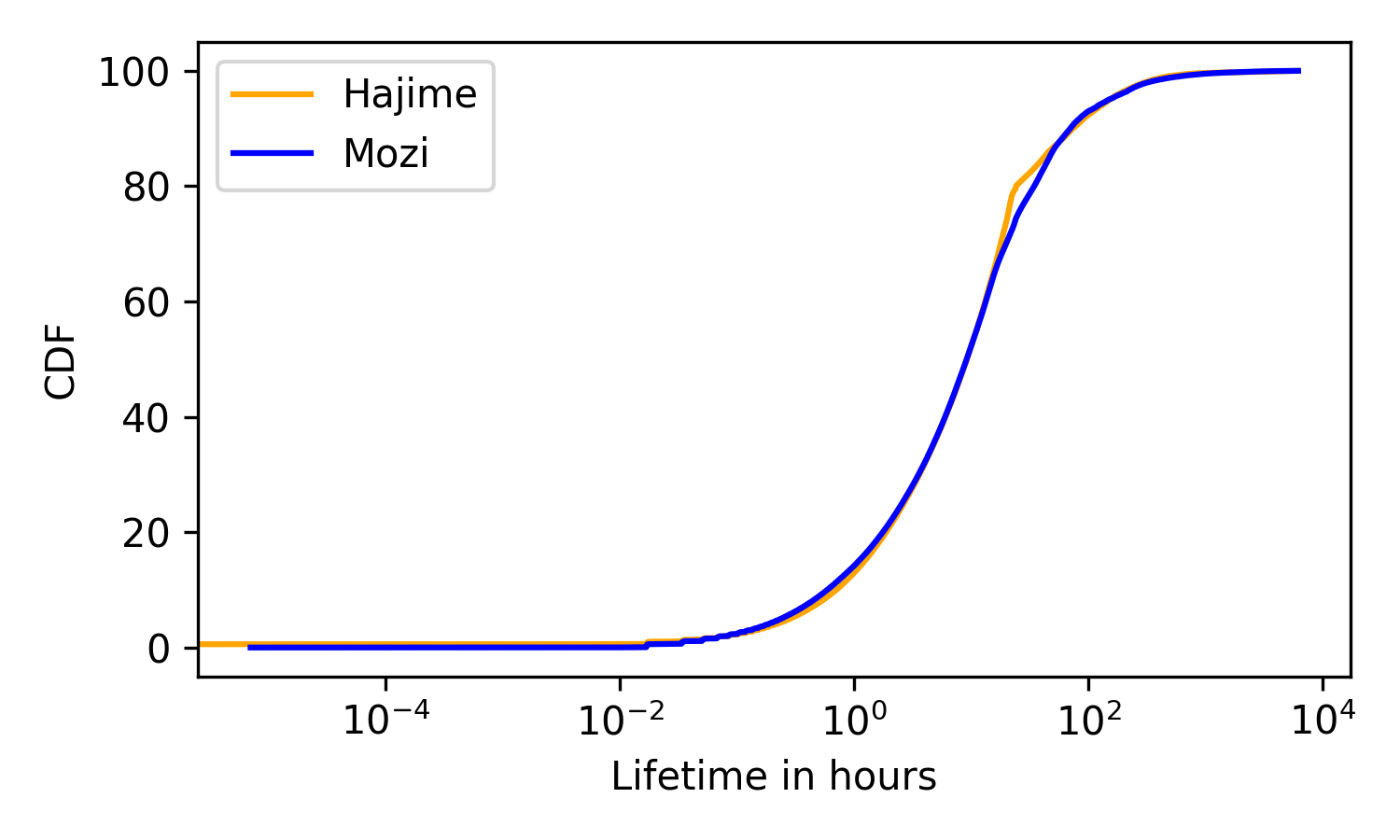}
    \caption{Cumulative distributions of bot lifetimes in Hajime and Mozi.}
    \label{fig:bot_lifetimes}
\end{figure}

\begin{figure}[h]
    \centering
    \includegraphics[scale=0.4]{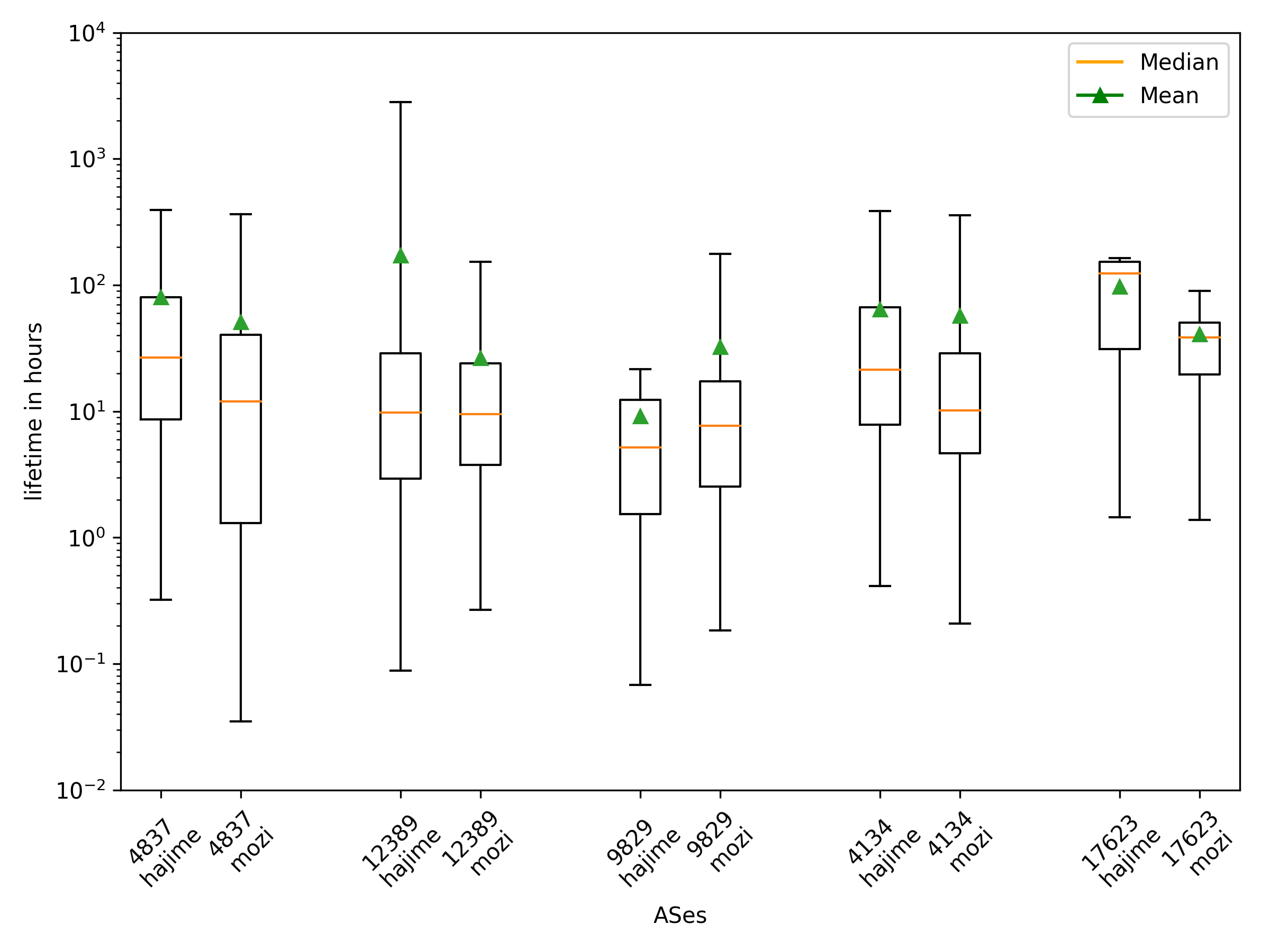}
    \caption{Boxplots of bot lifetimes in ASes that have a least 50 infections in Hajime and Mozi.} 
    \label{fig:lifetimes_asn}
\end{figure}

Within this section, we want to investigate the (dis-)similarity of bot lifetimes in Hajime and Mozi.
We measure a bot's lifetime as the duration between the first and last time we observed an ID in our measurements.
Hajime and Mozi change ID upon reboot or update and prevent us from tracking individual infections beyond a restart of the malware.
However, Mozi implements persistence mechanisms for most devices that re-infect a device after reboot. 
Since a re-infection after a reboot triggers a new ID generation, we cannot track Mozi infections across reboots.
That said, both botnets change ID for the same reason allowing a fair comparison of lifetimes between the two.

To obtain accurate lifetimes, we want to avoid counting two bots as one due to ID collisions.
As we could not distinguish ID collisions within an AS for Mozi, we filter all bots with a simultaneously active ID at two IP addresses for this analysis.

Figure~\ref{fig:bot_lifetimes} provides the cumulative distributions for the lifetimes of bots in Hajime and Mozi. 
We can see, that with minor deviations both botnets have similar lifetime distributions.
However, due to the wide range of the x scale, it is difficult to identify subtle yet important differences.
The most notable difference is that Mozi has a higher median lifetime of 7.32 hours vs. 5.55 hours in Hajime, whereas Hajime has a higher mean lifetime of 96.7 hours vs 41.13 hours in Mozi.
This is a percent difference between median lifetimes of 27.5\% and a percent difference between the mean lifetimes of 80.64\%.
Interestingly, the median lifetime of Hajime did not change much since 2018, when Herwig et al. reported a median lifetime of 5 hours~\cite{DBLP:conf/ndss/HerwigHHRL19}.
This discrepancy can largely be explained by Hajime having both more long and more short lifetimes than Mozi leading to a lower median yet higher mean lifetime.

While not extremely different, we want to understand what might cause the differences in lifetime.
From a naive point of view, one would expect devices like routers, security cameras and DVR's to be rarely turned off by their owners.
Therefore, we expect the most probable causes to be either the difference in demographics or a difference in the implementation of the botnet.
For this, we compare the lifetimes of all five ASes with at least 50 infections in both Hajime and Mozi.
Figure~\ref{fig:lifetimes_asn} shows the lifetimes for each AS in Hajime and Mozi.
We can see that the mean and median lifetimes differ for Hajime and Mozi even within the same AS.
The only exception is the median lifetime in AS 12389.
Overall the percent difference is 79.36\% for mean lifetimes and 58.84\% for median lifetimes. 
Furthermore, the lifetimes are longer for Hajime except for AS 9829.
These results indicate that lifetimes between botnets even differ in the same AS. 
Furthermore, we can see, that AS 17623 has much longer lifetimes than the other five ASes in the comparison.
Based on these findings, we speculate that the discrepancies in lifetimes are caused by a combination of affected ASes and the devices targeted by each botnet.

In summary, we found that the lifetime distributions of Hajime and Mozi follow similar trends with the majority of lifetimes between one and 100 hours.
Nevertheless, the mean and median lifetimes differ by 27.5\% and 80.64\% respectively, indicating that either the affected ASes or targeted devices have a non-negligible influence on a bots lifetime.

\subsection{Device Churn}

Another common measurement of botnets is device churn. 
Understanding the rate at which devices are infected and disinfected provides essential insights into the overall vulnerable population, bot decay, infection cleanups, and the addition of new exploits~\cite{DBLP:conf/ccs/GriffioenD20,DBLP:conf/ndss/HerwigHHRL19}.

\subsubsection{Global Population}

\begin{figure*}
    \centering
    \begin{subfigure}{0.48\textwidth}
        \includegraphics[scale=0.7]{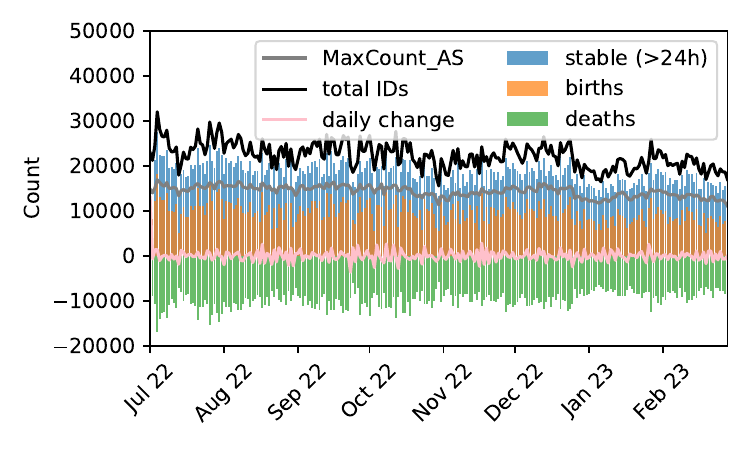}
    \end{subfigure}
    \begin{subfigure}{0.48\textwidth}
        \includegraphics[scale=0.7]{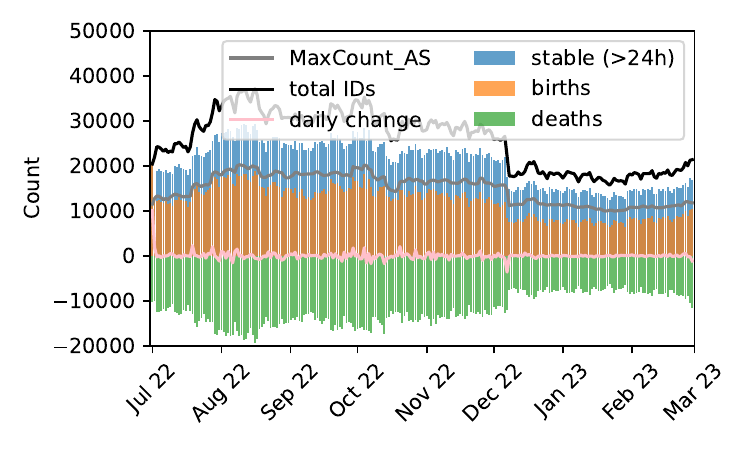}
    \end{subfigure}
    \caption{Births and deaths in Hajime (left) and Mozi (right).}
    \label{fig:size_change}
\end{figure*}

Figure~\ref{fig:size_change} provides a detailed look at the daily churn in Hajime and Mozi.
The Figure depicts daily births and deaths and their difference.
Furthermore, it shows the number of \emph{stable} IDs, i.e., those active at least 24 hours before the start of the current day, and the total number of IDs observed on a day.

Overall, Hajime and Mozi have a balanced ratio of births and deaths.
The balance of births and deaths indicates that the population size of both botnets is stable throughout the eight-month measurement. 
A balanced ratio of births and deaths has two possible explanations.
One, the rate at which new devices are infected cancels out the rate at which devices are disinfected.
Two, the infected devices remain the same, and the reboots cause births and deaths.
%
%% Both can "co-exist"
We want to point out that neither explanation is exclusive and we likely observe a mix of both.

%% Previous papers showed "spikes" upon addition of exploits etc. -- Mirai scanning is very quick especially for larger botnets (Griffioen)
Previous work investigating Hajime and Mirai reported that adding exploits or new username password combinations caused immediate spikes in the population~\cite{DBLP:conf/ccs/GriffioenD20,DBLP:conf/ndss/HerwigHHRL19}.
These spikes in population highlight that Mirai-based botnets are highly effective in identifying and infecting the vulnerable population.
Consequently, new infections are most likely either newly deployed devices or reinfections.
Since Hajime was not updated since 2018 and Mozi's botmasters were arrested in June 2021\footnote{https://mp.weixin.qq.com/s/Su0-uU5JaUrAh8ptTzTCsA}, we did not expect and did not observe any update-related spikes in population.
Therefore, it is unlikely that the addition of new exploits caused considerable growth for any of the two botnets.
New infections are most likely due to new vulnerable devices becoming reachable from the Internet.
This may either be caused by newly installed devices, or ISPs changing network configurations, e.g., changing customers from shared IP addresses behind a CG-NAT to individual IP addresses. 
Similarly, we would expect most disinfections to be caused by infected devices no longer being used or patched.
% avg churn hajime -11.541322314049587
% avg churn mozi -1.322314049586777
Throughout the eight-month measurement, the average daily population change is -11.5 for Hajime and -1.3 for Mozi, indicating a slight decline in size.

\begin{figure}
    \centering
    \includegraphics[scale=0.7]{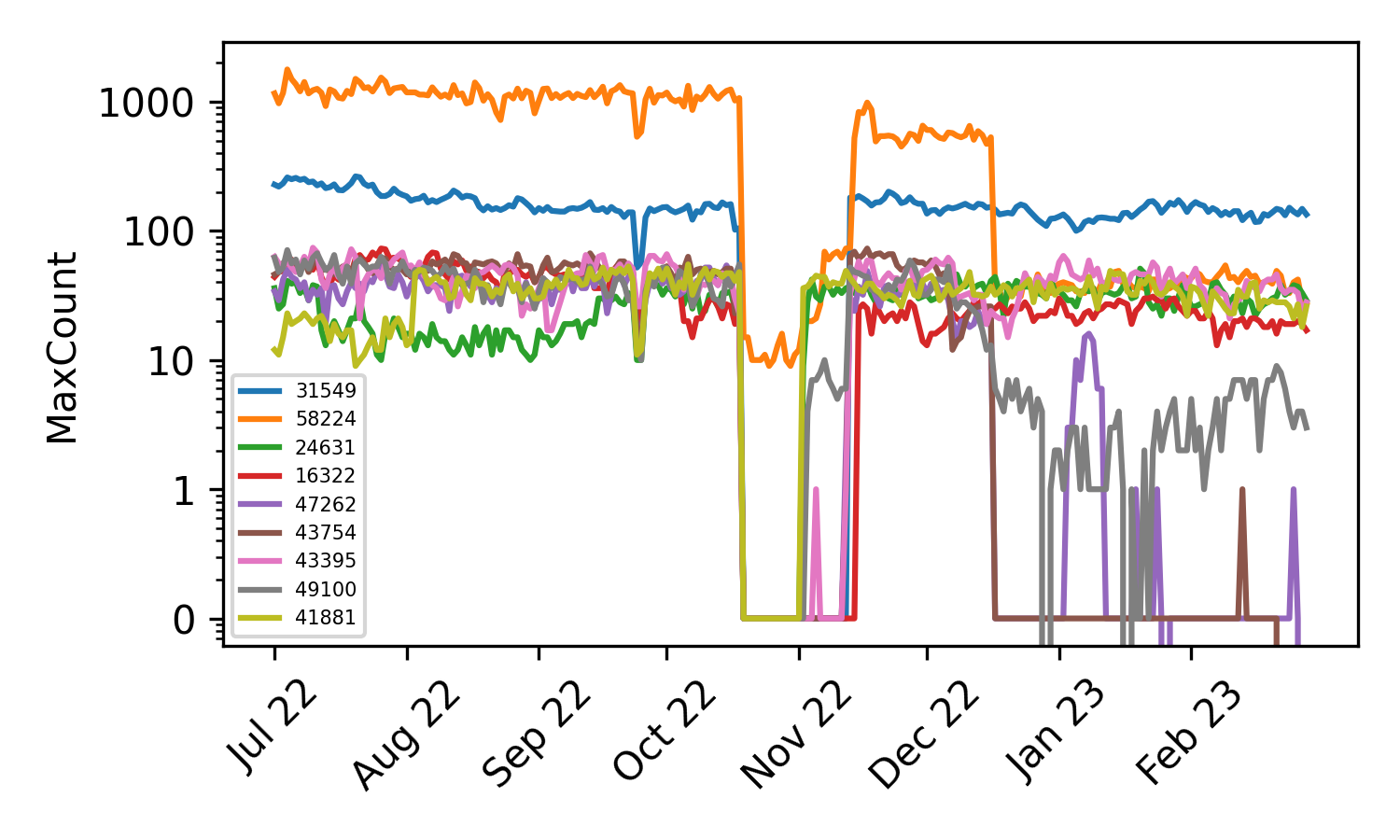}
    \caption{Internet Shutdown of Iranian ASes in October and November 2022}
    \label{fig:iranian_shutdown}
\end{figure}

Another interesting observation is the similar ratio of \emph{stable} and new IDs in both botnets.
Out of the average 22,249 IDs in Hajime, 9,822 IDs (44.15\%) were first seen on a given day, and 8,473 IDs (38.08\%) were active for more than 24 hours.
For Mozi, we see an average of 25,774 IDs daily, with 12,141 IDs (47.11\%) first seen and 8,457 IDs (32.81\%) active for more than 24 hours.
The remaining 17.77\% for Hajime and 20.08\% for Mozi are IDs that were first seen on the previous day.
In both botnets, the \emph{stable} IDs make up more than half of the average $MaxCount_{AS}$, indicating that most of the population is indeed stable and remains infected for multiple days.
Overall, given the stable population size, fast exploitation of vulnerable devices, and slight decrease in size indicate that device reboots primarily cause births and deaths.

While there is an overall slight variation in size, there are some drastic changes for both botnets.

While most of them are due to changes in individual ASes, which we discuss in the following section, two changes are caused by our measurement infrastructure.

% %
First, the growth in size for Mozi on the 19th of July 2022 is caused by adding the third crawler in a different geographic location.  
Investigating why this crawler improved our overall performance, we could not identify any reasons other than network location.
Second, the drop in the Hajime population between the 24th of December 2022 and the 26th of January 2023 is caused by a crawler malfunction.
Due to the holiday season, we were slow to identify an error that slowed down two of our Hajime crawlers.
% %
\subsubsection{AS Population}

\begin{figure*}
    \centering
    \begin{subfigure}{0.48\textwidth}
        \includegraphics[scale=0.7]{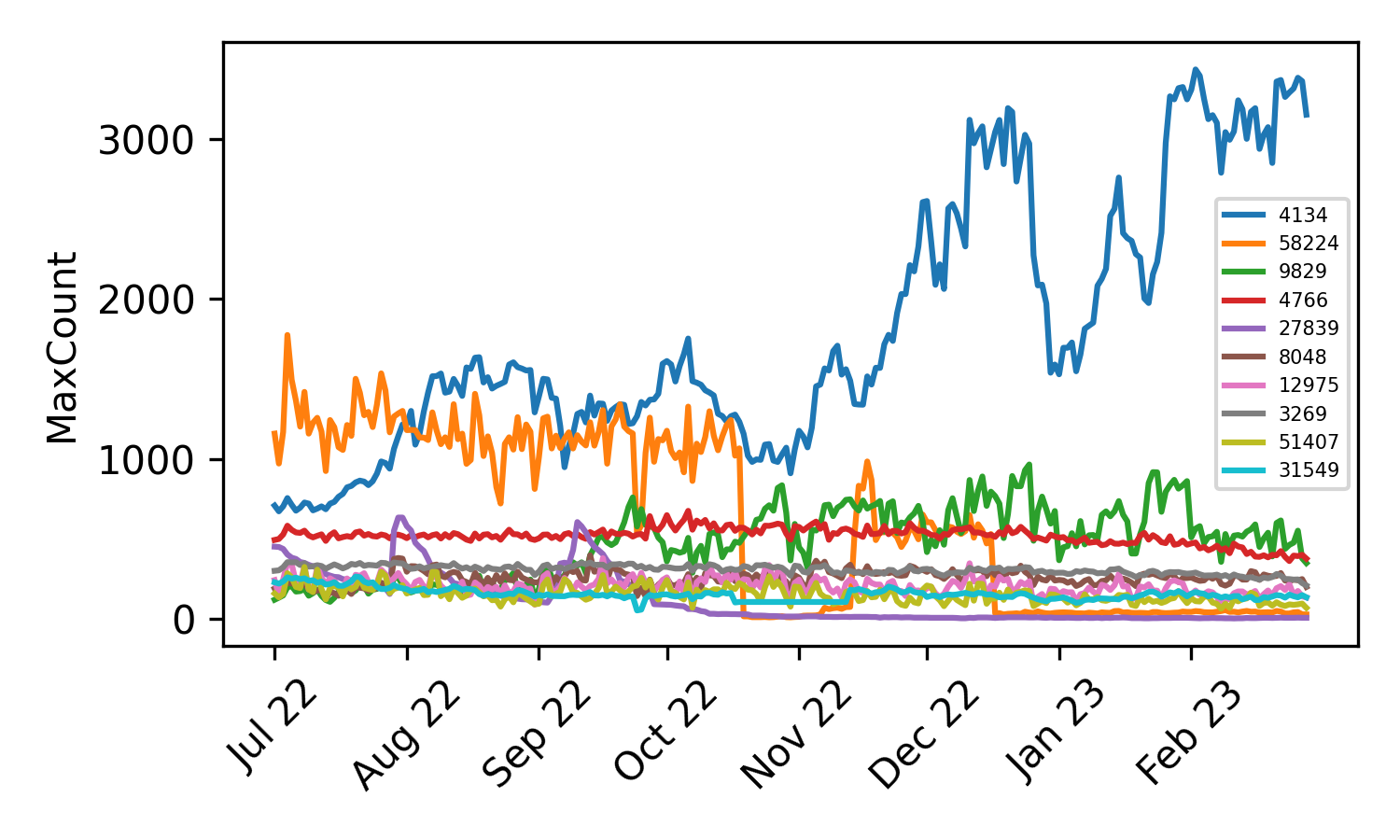}
    \end{subfigure}
    \begin{subfigure}{0.48\textwidth}
        \includegraphics[scale=0.7]{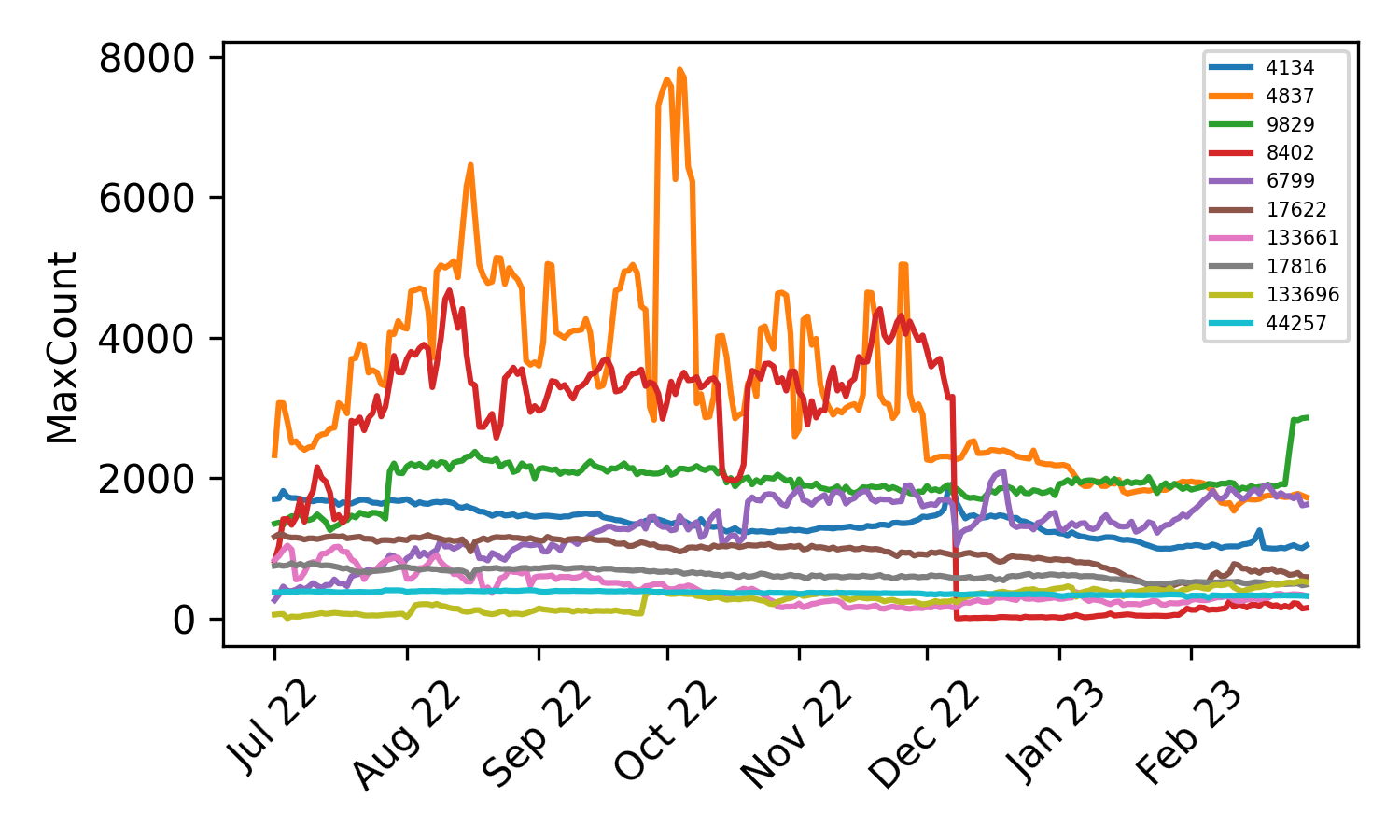}
    \end{subfigure}
    \caption{Size fluctuation for top 10 ASes in Hajime (left) and Mozi (right).}
    \label{fig:pop_change}
\end{figure*}

\begin{table}[]
    \centering
    \begin{tabular}{@{}lllrrr@{}}
    \toprule
    ASN    & Max  & Min  & \multicolumn{1}{l}{Mean} & \multicolumn{1}{l}{Median} & \multicolumn{1}{l}{STD} \\ \midrule
    \multicolumn{6}{c}{\textbf{Mozi}}                                                                      \\ \midrule
    4837   & 7817 & 1534 & 3,278.07                 & 3,009.00                   & 1,358.38                \\
    8402   & 4671 & 2    & 2,118.56                 & 2,914.00                   & 1,592.36                \\
    9829   & 2858 & 1260 & 1,935.44                 & 1,924.00                   & 256.14                  \\
    6799   & 2090 & 262  & 1,306.51                 & 1,317.00                   & 403.17                  \\
    4134   & 1918 & 994  & 1,363.44                 & 1,363.00                   & 216.01                  \\
    17622  & 1199 & 464  & 944.38                   & 1,009.00                   & 201.84                  \\
    133661 & 1040 & 139  & 414.67                   & 338.00                     & 228.03                  \\
    17816  & 791  & 472  & 627.35                   & 616.00                     & 77.52                   \\
    133696 & 533  & 4    & 262.18                   & 284.00                     & 143.68                  \\
    44257  & 404  & 296  & 363.00                   & 365.00                     & 24.41                   \\ \midrule
    \multicolumn{6}{c}{\textbf{Hajime}}                                                                    \\ \midrule
    4134   & 3436 & 672  & 1,806.53                 & 1,539.00                   & 796.98                  \\
    58224  & 1775 & 9    & 615.95                   & 554.00                     & 532.89                  \\
    9829   & 965  & 106  & 474.47                   & 502.00                     & 218.12                  \\
    4766   & 675  & 361  & 518.86                   & 526.00                     & 50.75                   \\
    27839  & 633  & 2    & 114.99                   & 15.00                      & 156.56                  \\
    8048   & 397  & 124  & 256.05                   & 258.00                     & 46.47                   \\
    12975  & 379  & 78   & 206.24                   & 204.00                     & 51.48                   \\
    3269   & 354  & 208  & 307.07                   & 310.00                     & 26.61                   \\
    51407  & 324  & 63   & 153.27                   & 149.00                     & 50.57                   \\
    31549  & 264  & 1    & 143.10                   & 150.00                     & 58.28                   \\ \bottomrule
    \end{tabular}
    \caption{Population Change for top 10 ASes.}
    \label{tab:pop_change}
\end{table}

Figure~\ref{fig:pop_change} and Table~\ref{tab:pop_change} provide detailed information about the size of the top 10 ASes by infection count over the measurement period.
In contrast to the relatively stable size of the global population, we can observe drastic changes for individual ASes.
Starting with Hajime, we can see that five ASes (51407, 12975, 3269, 8048, 4766) have only slight fluctuations in size. 
In contrast, ASes 4134, 9829, and 27839 show significant increases in size throughout the measurement, more than doubling in size. 
Furthermore, AS 27839 quickly decreases in size after two spikes and reaches nearly zero toward the end of the measurement.
Another interesting observation is, that the two Iranian ASes 58224 and 31549 both drop in size between October and November 2022. 
AS 58224 drops in size again in December 2022 and its size stays close to zero afterwards.
Figure~\ref{fig:iranian_shutdown} provides a closer look at all Iranian ASes in our dataset with at least ten infections. 
We can see, that almost all of the Iranian bots become unreachable at the same time. 
This countrywide drop in infections is most likely due to Internet censorship related to the Mahsa Amini protests\footnote{\url{https://en.wikipedia.org/wiki/Timeline_of_the_Mahsa_Amini_protests}}.
Lastly, the drop visible for AS 4134 and 9829, between the 24th of December 2022 and the 26th of January 2023 is the aforementioned crawler malfunction that slowed down two of our Hajime crawlers.

For Mozi, we can observe similar patterns, where some ASes have stable populations and others are growing or shrinking.
Interestingly, many of the more stable ASes (44257, 17816, 17622, 133696, 4134) visibly decrease throughout the measurement period, with some short peaks for AS 133661, and 4134.
This indicates a slow but noticeable decay of Mozi infections in those ASes, most likely caused by discontinued use of devices or non-coordinated remediation by the users.
However, we can also observe spiking increases in size for ASes 133696, 8302, 9829 and 4837. 
These types of spikes have previously been observed with the addition of new exploits~\cite{DBLP:conf/ndss/HerwigHHRL19,DBLP:conf/ccs/GriffioenD20}.
However, since the Mozi botmasters were arrested and no new exploits were added during the measurement, it is not related to new exploits.
Our best guess is, that bots previously infected by another botnet were disinfected and became vulnerable to re-infection, or changes in an ISP's network made a set of vulnerable devices routable from the Internet, causing a spike in infections.
For the latter, we found one example in our dataset, where a previously unseen subnet of AS 9829 correlated with the first spike in infections. 
For the other spikes, we did not observe a similar signal to draw further conclusions.
Lastly, we see a sudden drop in infections for AS 8402 on the 7th of December 2022.
We speculate, that this is related to a coordinated effort by the ISP, e.g., quarantining infected devices~\cite{DBLP:conf/ndss/CetinGAKITTYE19} or a coordinated patch of ISP-controlled routers.
Interestingly, we do not see a drop in the ~20 Hajime bots in AS 8402 at all, indicating that this event was indeed targeted at Mozi.

To us, the most interesting finding is that in both Hajime and Mozi we see drastic increases in size for some ASes.
As we mentioned previously, this type of event is similar to the addition of new exploits, which we can rule out for both botnets.
A possible explanation for this is our observation that many infections share IP addresses, indicating the presence of a CG-NAT. 
This is specifically the case for ASes with large spikes in infection count.
At this point, we can only speculate how these devices get infected in the first place, as the vulnerable ports should not be directly reachable from the Internet.
However, if a single device within a CG-NAT it might be able to infect all other vulnerable devices in that network.
Since we could observe a similar pattern for AS 9829 we believe this to be the most plausible explanation for the sudden increases in bot infections.

In summary, we found that Hajime and Mozi follow the same trends when it comes to bot lifetimes and device churn.
Nevertheless, bot lifetimes in Hajime are on average more than twice as long as bot lifetimes in Mozi.
Contrary, median lifetimes in Mozi are 27.5\% longer than in Hajime, showing subtle yet important differences between the two botnets.
Further investigations on an AS level have shown, that ASes can differ largely both in bot lifetimes and bot churn.
Furthermore, we did not only find differences across ASes but also within the same AS for different botnets.
Two examples of this are more than twice as long bot lifetimes for Hajime in AS 17623 and frequent changes in growth for AS 4134 in Hajime, but a mostly stable population in Mozi.
Therefore, while we can expect that other IoT botnets will largely follow the same trends concerning bot lifetime and churn, we need to expect differences if the AS composition is largely different from Hajime and Mozi.

%% file: replyrate.tex
\section{Reply Ratio}

As explained before, both Hajime and Mozi built their command and control channel on top of the Bittorrent DHT. 
However, they each use a different approach to retrieve the latest configurations.
Within this section, we want to compare the effectiveness of retrieving configuration files in each botnet.
To do this, we measure the probability of receiving a configuration file upon sending a request as follows:

$$ \mbox{reply ratio} = \frac{\mbox{requests sent}}{\mbox{configurations received}} $$ 

\begin{figure}
    \centering
    \includegraphics[scale=0.50]{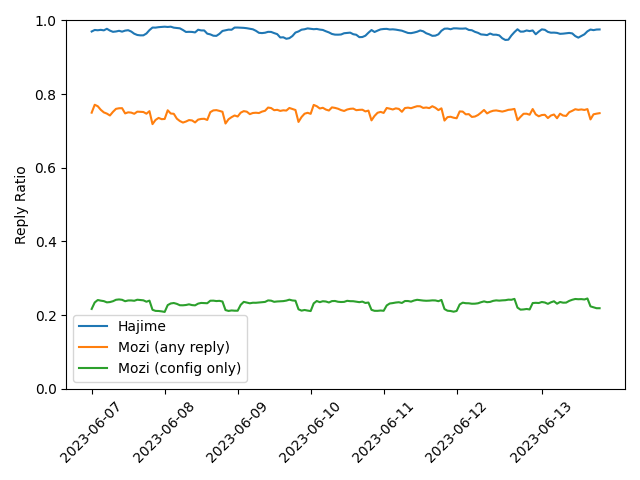}
    \caption{Reply ratios for Hajime and Mozi botnets. }
    \label{fig:fail_rate}
\end{figure}

Figure~\ref{fig:fail_rate} depicts the reply ratio for Hajime and Mozi\footnote{Due to an oversight in recording \emph{find node} replies as neither a timeout nor a valid configuration, this evaluation was repeated on measurements in June 2023.}.
As expected, the overall reply ratio for Mozi is much lower, as it only sends a config with a probability of 1/3 upon receiving a \emph{find node} request.
To account for this difference in implementation, we also plotted the reply ratio if any reply from a Mozi bot is considered.
Interestingly, the reply ratio for Mozi on average is still much lower than for Hajime.
This indicates, that Mozi bots are less reliable to respond than Hajime bots. 
A possible explanation for this could be computationally weak devices~\cite{DBLP:conf/uss/AntonakakisABBB17}, causing a higher packet loss.
Furthermore, we observe diurnal drops in reply rate, which as we will show later are largely related to bots in Chinese ASes.
Based on the significantly worse replay ratio of Mozi, we deem the approach implemented by Mozi to be inferior to Hajime.
In practice, the dissemination of configurations will either be slower or more likely to be noticed due to repeated requests.
Moreover, if Mozi would switch to returning configurations 100\% of the time, they would fully rely on non-bots to discover peers, as bots would never reply with \emph{find node} responses.

What makes this discrepancy even more interesting is, that Hajime requires an additional round trip of messages to retrieve the uTP and config from a bot.
Therefore, the likelihood of a timeout or other failure should be more likely than for Mozi.

\begin{figure}
    \centering
    \includegraphics[scale=0.50]{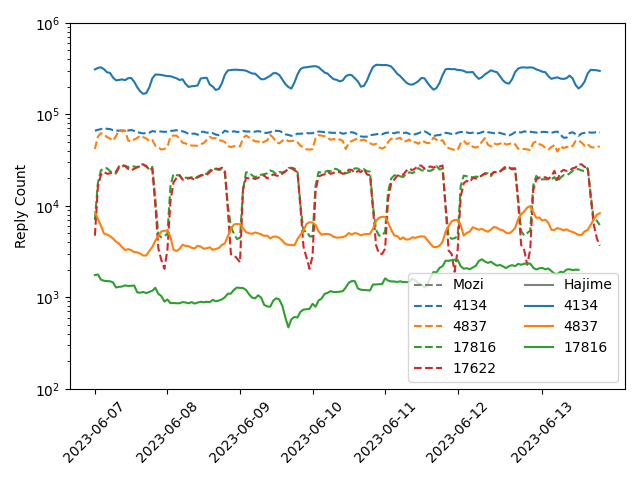}
    \caption{Reply count for ASes in China.}
    \label{fig:reply_rate_china}
\end{figure}

\begin{figure*}
    \centering
    \begin{subfigure}{0.48\textwidth}
        \includegraphics[scale=0.55]{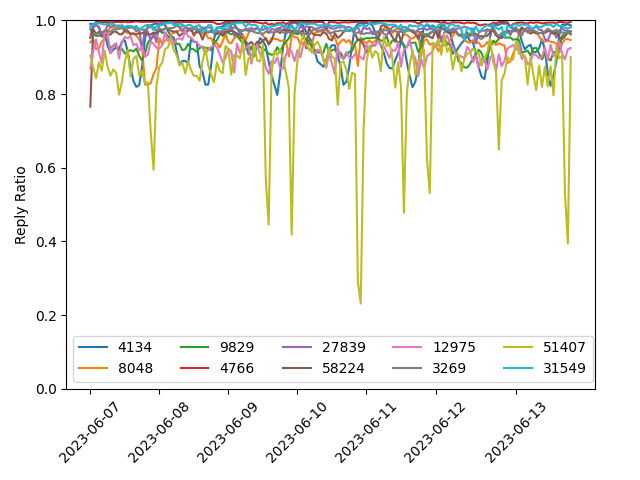}
        % \caption{Hajime}
        % \label{fig:mozi_size}
    \end{subfigure}
    \begin{subfigure}{0.48\textwidth}
        \includegraphics[scale=0.55]{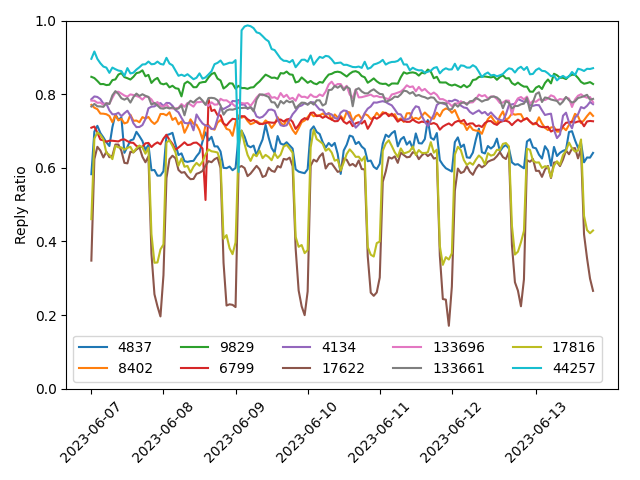}
        % \caption{Mozi}
        % \label{fig:hajime_size}
    \end{subfigure}
    \caption{Reply ratio for top 10 ASes in Hajime (left) and Mozi (right).}
    \label{fig:fail_rate_as}
\end{figure*}

In addition to the overall reply rate, we want to once again look at differences on an AS level. 
Figure~\ref{fig:fail_rate_as} depicts the reply ratio for the top 10 ASes in each botnet.
We can see in both botnets that the AS influences the reply ratio.
This becomes apparent by some ASes having a consistently lower reply ratio than others.
Generally, the difference between ASes is larger for Mozi.
Furthermore, we can see that some ASes have extreme drops in their reply ratio. 
While most of them appear to be diurnal, this is not the case for AS 51407 in Hajime.
We suspect that the short, irregular, drops may be related to the ongoing Israeli-Palestinian conflict\footnote{\url{https://en.wikipedia.org/wiki/Timeline_of_the_Israeli\%E2\%80\%93Palestinian_conflict_in_2023}}.
Furthermore, we find that the diurnal patterns in ASes (4134, 4837, 17622, 17816) are all located in China. 
Figure~\ref{fig:reply_rate_china} provides a closer look at these ASes for both botnets.
We were expecting, that the diurnal patterns are related to the networks themselves. 
As an example, Zhu et al. reported, that transnational traffic is slowed down during certain times of day~\cite{DBLP:conf/sigmetrics/ZhuM0QEHD20}.
The reported window measured by them aligns closely with the drop in reply ratio for AS 17622 and 17816 in Mozi.
However, we do not see the same drop for AS 17816 in Hajime, raising concerns that traffic throttling is the actual cause.
Furthermore, AS 4134 and 4837 in Hajime even peak at that time of day.
Generally, none of the ASes for which we have infections in Hajime and Mozi present a similar pattern in reply rate. 
Consequently, while there seems to be a trend towards diurnal patterns in China, the cause can not be generalized to patterns within ASes or the country as a whole.

To summarize, Hajime and Mozi each implement a peer-to-peer command and control channel on top of the Bittorrent DHT.
While both implementations provide the desired functionality, Hajime's approach appears much more efficient.
Specifically, if we consider that Mozi bots only share their configuration at a probability of 1/3, a Hajime bot is four times more likely to retrieve a configuration file upon requesting it from another bot.
Furthermore, to overcome this disadvantage Mozi would have to increase its communication frequency, raising the likelihood to be detected. 
Nevertheless, both peer-to-peer architectures have proven to be highly resilient, as the two botnets remain active whereas many centralized Mirai variants have been taken down.
Most notably, Mozi remains active even after the arrest of the botmasters and even manages to infect new devices.

%% file: discussion.tex
\section{Discussion}
\label{sec:discussion}

In this paper, we have analyzed how Hajime and Mozi have evolved over the common ancestor Mirai. 
We find, that the addition of new exploits caused the infections of the two botnets largely separated not just in terms of targeted hosts but even in the targeted countries and ASes.
Furthermore, even though both botnets adapt more resilient peer-to-peer based command and control channels, the implementations and reliability of retrieving configuration files are different.
We also found that, while following similar distributions, the lifetimes of bots differ greatly in both botnets.
We could identify some of these differences to be rooted in the devices or ASes targeted by each botnet.
Therefore, considering that even though both botnets target similar devices, e.g., routers, cameras and DVR's, their populations and characteristics are unique.
While the early Mirai variants essentially competed over the same pool of devices~\cite{DBLP:conf/ccs/GriffioenD20}, these latest variants have become mostly separate.
Therefore, we want to argue that the \emph{canonical IoT botnet} does not exist anymore, and new IoT botnets should be carefully studied to identify and understand their unique characteristics.

Another topic of concern is, that in some ASes both Hajime and Mozi were able to infect large numbers of bots without the addition of new exploits.
We speculate that this is related to CG-NAT deployments.
While vulnerable devices behind a CG-NAT should theoretically not be affected by scanning vulnerable ports, we see infections of devices in CG-NATs in our dataset.
A likely explanation is that once a \emph{patient zero} gets infected within a CG-NAT, it can then attack other vulnerable devices.
If this is indeed true, a botnet capable of effectively bypassing CG-NAT, e.g., by adding attack vectors through email, could vastly exceed the size of IoT botnets seen today.
Specifically, if we consider that currently targeted devices like routers, cameras, and DVRs are the only routable but nowhere near the only IoT devices deployed in many households.
Hence, while many Mirai infections have been remediated, there remains the threat of an \emph{uber botnet}, that not only combines the populations of Haime and Mozi but finds a way to effectively infect IoT devices within (CG-)NAT networks.

%% file: conclusion.tex
\section{Conclusion}
\label{sec:conclusion}

Due to its powerful attacks and publicly available code, Mirai
established itself as the canonical IoT botnet early in the burgeoning
days of IoT.
Griffioen and Doerr~\cite{DBLP:conf/ccs/GriffioenD20} observed many
Mirai variants, and found that they competed with one another for many
of the same IP addresses.
They also found that the most successful variants were those that
evolved to include novel vulnerabilities, defenses, and regional
concentrations.
Nonetheless, despite these minor differences at the time, there was
enough overlap between Mirai variants to indicate that Mirai was
\emph{the} canonical IoT botnet.

In this paper, we revisited the battle for IoT botnet supremacy by
measuring two of the most popular Mirai descendants: Hajime and Mozi.
To actively scan both botnets simultaneously, we developed a robust
measurement infrastructure, \name, and ran it for more than eight
months.
The resulting datasets show that these two popular botnets have
diverged in their evolutions from their common ancestor.
For instance, while Griffioen and Doerr found that Mirai variants used
to compete extensively over individual IP
addresses~\cite{DBLP:conf/ccs/GriffioenD20}, we find that there is
nearly zero overlap today.
Moreover, our results showed that the two botnets differ in how they
react to external events, such as China's diurnal rate-limiting of
transnational connections~\cite{DBLP:conf/sigmetrics/ZhuM0QEHD20}.

Collectively, our results show that there is no longer one canonical
IoT botnet.
We have entered an era in which the properties of one IoT botnet
are not likely to generalize to those of another.
This conclusion has several important implications.
\emph{First}, it means that understanding IoT botnets writ large
requires investigating multiple botnets simultaneously.
\emph{Second}, it means that defending against IoT botnets
will require a diverse set of tools; what works to detect or
mitigate one botnet may not work for another, as they exhibit
different strengths and weaknesses.
\emph{Finally}, it means that there is the potential for another
phase of evolution that seeks to combine the disparate botnets
into a ``grand unified botnet.''
Because the various Mirai botnets have, over time, added their
own unique sets of vulnerabilities, such a combined botnet 
could be larger than previous Mirai instantiations.
To assist in efforts in measuring and identifying future such threats,
we will be making our measurement infrastructure, \name, publicly
available.
%